\documentclass[aps,pra,preprint]{revtex4-1}
\usepackage{amssymb,amsmath}
\usepackage{graphicx}
\usepackage{grffile}
\usepackage{mathtools}
\DeclarePairedDelimiter\floor{\lfloor}{\rfloor}
\begin{document}

\title{Rates of convergence of the partial-wave expansion beyond Kato's cusp condition II: evaluations for the prefactors on the ground state of the helium atom}
\author{Cong Wang}
\email[]{wangcon9@msu.edu}
\affiliation{Department of Chemistry, Michigan State University,  East Lansing, Michigan 48824, USA }

\date{\today}

\begin{abstract}
This article is a continuation of our previous work (Phys. Rev. A 88, 032511 (2013)). The prefactors for the partial-wave expansion of the helium atom are derived.
Due to series of cancellations, the partial-wave increments of the energy converge as $L^{-2N-6}$.
The origin of these cancellations is identified from alternative expressions of the partial-wave energies.
 There is some evidence that the assumptions of the regularities for the exact wavefunction can be reduced.

\end{abstract}

\maketitle

\section{Introduction}

One goal of electronic structure theory is to solve the eigenvalue problem of a many-electron Hamiltonian
\begin{equation}
H | \Psi \rangle = E | \Psi \rangle \label{1}
\end{equation}
The general strategy is to expand the eigenstate into an $N$-electron basis and evaluate the expansion coefficients.
Since the closed-form exact solution is not available, the rates of convergence with respect to a basis would be valuable. \\

There are two aspects to the rates of convergence \cite{Nakatsuji:12a}. One is related to the hierarchy of correlations inside the exponentially growing Fock space.
The other is the functional form of the one- and two-electron bases.
The former issue arises from the fact that the eigenstate of an additive Hamiltonian is an antisymmetric product of subsystems' states. When interactions between subsystems are added, the expansion amplitudes on all bases will be changed.
The computational difficulty could be viewed as a consequence of using a classical computer to simulate the quantum system. It may be solved by quantum computing \cite{Feynman:82,Alan:05,Troyer:14}. Fortunately, all interactions in the standard model are local and we are primarily interested in phenomena with rather limited energy scales. Hence, various truncation schemes, such as coupled cluster  \cite{Coester:58,Coester:60,Bartlett:07,lyakh2012multireference} and density-matrix renormalization group \cite{White:92,Head-Gordon:03} provide useful results on classical computers. The latter issue is related to the local and global properties of the exact wavefunction \cite{Kato:57,Gilbert:63,Geer:01,Fournais:05,Nakatsuji:07,Fournais:09}. It could be improved by including the interelectron distance into the basis.

The most straightforward construction of an $N$-electron basis is an antisymmetrized product of one-electron spin orbitals.
One prototype is the partial-wave expansion (PWE) of the ground-state of the helium atom \cite{Kutzelnigg:92}
\begin{align}
\psi (r_1,r_2,r_{12} ) &= \sum_{l=0}^L \psi_l(r_1,r_2) P_l (\cos \theta_{12})
\end{align}
where $P_l(\cos \theta_{12})$ is the Legendre polynomial. We have omitted the spin functions and assumed the wavefunction is real throughout. Here $\psi_l$ can be defined in at least two ways. The first is an expansion for the exact, either the first-order perturbative wavefunction or the eigenfunction of the full Hamiltonian, \textit{i.e.}, $ \psi_l := \frac{2l+1}{2} \int_0^{\pi} \psi_{\mathrm{exact}} P_l(\cos \theta_{12} ) \sin \theta_{12} d\theta_{12} $.
The second  is the wavefunction corresponding to the optimized energy, either via the Hylleraas functional of the second-order energy or the Ritz's procedure of the non-perturbative energy. We shall denote the optimized quantities as $\tilde{\psi}_l$ and $\tilde{E}(L)$.

It was shown by Schwartz \cite{Schwartz:62} that at the large-$l$ limit, the asymptotic first-order wavefunction in the $1/Z$ expansion is
\begin{equation}
\psi|_{r_1=r_2} \rightarrow \frac{1}{2} r_{12} \Phi|_{r_1=r_2} \label{Sl}
\end{equation}
where $\Phi = e^{-r_1 - r_2} /\pi$ is the unperturbed  wavefunction. The asymptotic expression here means that it provides the large-$L$ expression of the second-order energy, $E_2(L)$.  Schwartz then used a series expansion in terms of $(r_1-r_2)/(r_1+r_2)$ at $r_{12}=0$ for the first-order wavefunction to obtain the large-$L$ expansion of the second-order energy \cite{Schwartz:63}
\begin{equation}
 E_2(L)-E_2(L-1) = - \frac{45}{256} (L+1/2)^{-4}  + \frac{225}{1024} (L+1/2)^{-6} + \cdots \label{2ptpwe}\\
\end{equation}

Similar rates of convergence were found in Ritz's variation  \cite{Hill:85,Goddard:09b} and in many-electron systems \cite{Klopper:99,Jankowski:06,Klopper:12,Valeev:12,Ten-no:12,Hirata:12,Schwenke:12,Shepherd:12}.
Among these works, one remarkable result of Hill's formulation \cite{Hill:85} is, the bound of the difference between the optimized and expanded energies was derived:
$|\tilde{E}(L) - E(L)| \leq O(L^{-5})$.

On the perturbative side, Eq. (\ref{Sl}) was further interpreted by Kutzelnigg and Morgan (KM) \cite{Kutzelnigg:92}. It is claimed that the first-order wavefunction can be written as
\begin{equation}
\psi = \frac{1}{2} r_{12} \Phi  + O(r_{12}^2) \label{KM}
\end{equation}

\noindent as the cusp condition at the two-electron-coalesces region
\begin{equation}
\left. \frac{ \partial \psi}{ \partial r_{12}} \right|_{r_{12}=0} = \left. \frac{1}{2} \Phi \right|_{  r_{12}=0} \label{cuspKM}
\end{equation}
The PWE convergence (\ref{2ptpwe}) was derived from Eq. (\ref{KM}) including $O(r_{12}^3)$. Eq. (\ref{cuspKM}) is a special form of Kato's cusp condition for the first-order wavefunction  \cite{Kato:57,Kutzelnigg:92}.

Introducing the cusp condition into the many-electron wavefunction was known as the R12 approach \cite{Klopper:12,Valeev:12,Ten-no:12}. For the helium atom, the ansatz
\begin{equation}
\psi  =  \frac{1}{2} r_{12} \Phi + \sum_{l=0}^L \psi_l(r_1,r_2) P_l (\cos \theta) \label{ansatz1}
\end{equation}

\noindent leads to $L^{-8}$ rate of convergence of the increment of the second-order energy.  Ten-no \cite{Ten-no:04} introduced a Slater-type correlation factor as the F12 method. The asymptotic behaviors of the correlation factors at the large-$r_{12}$ limit have been discussed by Lesiuk, Jeziorski, and Moszynski \cite{Jeziorski:13,:/content/aip/journal/jcp/142/12/10.1063/1.4915272}. K{\"o}hn generalized the reference function $\Phi$ beyond the Hartree-Fock wavefunction as the extended SP (XSP) ansatz \cite{Kohn:10}. Nowadays the F12 ansatz has been combined with perturbation, coupled-cluster, and canonical transformation approaches as one of the standard methods for quantum many-electron systems \cite{Hirata:08,Klopper:12,Valeev:12,Ten-no:12,Hirata:12,Yanai:12}. The high-order cusp conditions were derived by Rassolov and Chipman \cite{Rassolov:96},  Tew \cite{Tew:08}, and Kurokawa, Nakashima, and Nakatsuji \cite{Nakatsuji:13,Nakatsuji:14}.

Besides the above developments of explicit correlation methods, there is a subtle issue in Eq. (\ref{KM}). Both Schwartz's result (\ref{Sl}) and the cusp condition (\ref{cuspKM}) are valid at $r_1=r_2$, while the behavior (\ref{KM}) contains information of $\Phi$ at the off-coalescence region, \textit{i.e.}, $r_1 \neq r_2$. Furthermore, the wavefunction at $r_1 \neq r_2$, in terms of $\partial \Phi / \partial r_<|_{r_<=r_>}$ and $\partial^2 \Phi / \partial r_<^2|_{r_<=r_>}$, was used by KM to obtain the prefactor at $O(L^{-6})$ in  the rates of convergence \cite{Kutzelnigg:92}. In the previous work \cite{Wang:13}, an attempt was made to derive Eq. (\ref{cuspKM}) from a series expansion of the exact first-order wavefunction. As shown in the erratum of the previous work \cite{Wang:15a}, such a derivation is not valid. As shown in Section II in the present work, any term proportional to $(r_1-r_2)^2 e^{-r_1-r_2}$ will lead to the same $L^{-6}$ rate of convergence in Eq. (\ref{2ptpwe}). In the present article, we shall discuss this point in evaluating the prefactors for the inverse-power law of the PWE.

The present work is organized as follows. In Section \ref{sec2a}, the previous formulation is recapitulated. In Section \ref{sec2b}, an alternative expression for the PWE energy is derived. It leads to $L^{-2N-6}$ rate of convergence. This implies several terms at $L^{-N-7}$ must cancel each other to be consistent. In Section \ref{sec2c}, White and Brown's formal solution of the first-order wavefunction \cite{White:67a,White:67c,White:67d} is introduced. The relation between the formal solution and the large-$l$ behavior (\ref{KM}) is discussed. In Section \ref{sec2d} and \ref{sec2e}, the prefactors of the PWE energies are computed using the formal solution without and with the geminal functions, respectively. The cancellation at $L^{-N-7}$ is verified. The off-coalescence region information in Eq. (\ref{KM}) is clarified.

In Section \ref{sec3}, the derived rates of convergence including prefactors are compared with numerical results. In Section \ref{sec4}, the derivation is extended to the Ritz's variational formulation. In Section \ref{sec5}, the variational rates of convergence are compared with the numerical results.

\section{Rates of convergence for the second-order energy of the $1/Z$ expansion}
\subsection{Recapitulation of the previous expressions: construction of the PWE with geminal functions} \label{sec2a}

The general scheme of the previous work \cite{Wang:13} is to construct a power series of $r_{12}$ for the exact wavefunction. Leading odd-power $r_{12}$ terms are set into the trivial wavefunction. The rest of the exact wavefunction regard as the PWE.

We consider the Hylleraas functional  \cite{Hylleraas:30} for the second-order energy
\begin{equation}
E_2(L):= \langle \psi^{(a)} | H_0 - E_0 | \psi^{(a)} \rangle + 2 \langle \psi^{(a)} | H' - E_1 |\Phi \rangle \label{Hy}
\end{equation}

\noindent where $\Phi$ is the eigenfunction of $H_0$. The exact first-order wavefunction $\psi$ has been decomposed into $\psi^{(a)}$, which is used to evaluate the second-order energy, and a remainder $\psi^{(b)}$.
\begin{equation}
\psi= \psi^{(a)}  + \psi^{(b)} \label{a1}
\end{equation}
Wavefunction $\psi^{(a)}$ consists of a geminal function $\psi_g$ and a partial-wave function $\chi$
\begin{equation}
\psi^{(a)}  = \psi_g  +  \chi
\end{equation}
\begin{equation}
\psi_g  =  \sum_{n=1,3,5,\cdots}^{N} r_{12}^n \Phi_n (r_1,r_2) , \,\,\,\,\,\, \chi = \sum_{l=0}^{L} P_l (\cos \theta_{12}) \chi_l (r_1,r_2) \label{a2}
\end{equation}
where $N$ is an odd number for the upper bound of the power of $r_{12}$. $\Phi_n$ is defined via a series expansion of $r_{12}$ for the first-order wavefunction $\psi$ at its minimum value $|r_1 -r_2|$
\begin{equation}
\psi (r_1,r_2,r_{12}) = \sum_{m} \left. \frac{ \partial^m \psi}{\partial r_{12}^m} \right|_{r_{12} = |r_1 - r_2|}  \frac{ \left(r_{12} - |r_1 - r_2| \right)^m }{m!} \label{taylor}
\end{equation}
such that
\begin{align}
\Phi_n &= \frac{1}{n!} \left[ \sum_{m} \frac{ (-|r_1 - r_2|)^m }{m! \,\,} \left. \frac{ \partial^{n+m} \psi }{\partial r_{12}^{n+m} }\right|_{\substack{ r_{12}= |r_1 - r_2|}} \right]  = \frac{1}{n!} \left[ \sum_m \left. \frac{ (r_<-r_>)^m}{m!} \frac{ \partial^{n+m} \psi }{\partial r_<^n r_{12}^m} \right|_{ \substack{ \hspace{0.1cm} r_<=r_>, \\ \hspace{-0.2cm} r_{12}=0} }  \right]  \label{Phi}
\end{align}
$\chi_l$  is the PWE of the first-order wavefunction substracted by the odd-power geminal functions
\begin{align}
\chi_l &= \frac{2l+1}{2} \int_0^{\pi} \big[ \, \psi (r_1, r_2, r_{12}) - \sum_{n=1,3,5,\cdots}^{N} r_{12}^n \Phi_n(r_1, r_2) \, \big] P_l (\cos \theta_{12}) \sin \theta_{12} d \theta_{12} \label{af}
\end{align}

Hence, $\psi^{(b)}$ is the remaining components of the PWE
\begin{equation}
 \psi^{(b)} = \sum_{l={L+1}}^{\infty} P_l (\cos \theta_{12}) \chi_l (r_1,r_2)
\end{equation}
Since Eq. (\ref{Phi}) includes all power of $r_1$ and $r_2$, $\chi_l$ corresponds to the Sack expansion \cite{Sack:64,Cohl:13} of $r_{12}^{N+2}$ for the leading-order rate of convergence. Namely,
\begin{align}
r_{12}^{\nu} &= \sum_{l=0}^{L_1} R_{\nu l} (r_1,r_2) P_l (\cos \theta_{12}) \\
R_{\nu l} (r_1,r_2)  &= \sum_k^{L_2} C_{\nu l k} r_<^{l+2k} r_>^{\nu-l-2k} \\
C_{\nu l k } &= \frac{ ( -\nu/2)_l }{ (1/2)_l } \frac{ (l - \nu/2)_k }{ (l+3/2)_k }  \frac{ (-1/2-\nu/2)_k }{k!}
\end{align}
where $L_1 = \infty$ for odd $\nu$. $L_2 = \floor*{ (\nu+1)/2 }$ is defined with the floor function. $x_n:= x (x+1) \cdots (x+n-1), n\geq 1 $, as the the Pochhammer symbol.

For example, when $N=1$, the geminal function includes $r_{12}, r_{12} (r_1 - r_2)^2,r_{12} (r_1+r_2) (r_1 - r_2)^2 \cdots$ terms and $\chi_l$ relates to the Sack expansion of $r_{12}^3$. No odd power of ($r_1$ - $r_2$) would appear due to symmetry of the ground state \cite{Nakatsuji:07}.

Additionally, if we reduce the upper bound of $m$ to $N-n+1$ in Eqs. (\ref{taylor}) and (\ref{Phi}), $\chi_l$ will include the Sack expansion of $r_{12} (r_1 - r_2)^2$ and $r_{12}^3$. The inverse power of $L$ in the rates of convergence remains the same, but with a different prefactor. This possibility will be listed in the supplementary material.

\subsection{Alternative expressions of the second-order energies} \label{sec2b}

In the previous work, $E_2(L)$ was decomposed as
\begin{align}
E_2(L) &= A_2 + I_2(L) \\
A_2 &:= \langle \psi_g | H_0 - E_0 | \psi_g \rangle + 2 \langle \psi_g | H' - E_1 | \Phi \rangle \\
I_2 (L) &:= \langle \chi | H_0 - E_0 | \psi_g \rangle + \langle \psi_g | H_0 - E_0 | \chi \rangle +  \langle \chi | H_0 - E_0 | \chi \rangle  +  2 \langle \chi | H' - E_1 | \Phi \rangle
\end{align}

\noindent for the non-PWE and PWE parts, $A_2$ and $I_2(L)$, respectively. With the help of the first-order equation,  $I_2(L)$ was written as \cite{Wang:13}
\begin{align}
I_2(L) &=  \sum_{m=1}^{N} \langle \chi | -m(m+1)r_{12}^{m-2} + U_m + r_{12}^m (H_0 - E_0 )  | \Phi_m \rangle +  \langle \chi | H' - E_1 | \Phi \rangle \label{p1}
\end{align}

\noindent where the operator $U_m$ was introduced by $[ H_0 - E_0, r_{12}^m \,] = - m(m+1) r_{12}^{m-2} + U_m$. The leading terms in (\ref{p1}), $\langle \chi | U_m | \Phi_m \rangle |_{m=1} $ and $-m(m+1) \langle \chi | r_{12}^{m-2} | \Phi_m \rangle|_{m=1}$, converge as $L^{-N-7}$. It was then concluded that $E_2(L) \rightarrow L^{-N-7}$ for a sufficient large $L$ \cite{Wang:13}.

However, the Hylleraas functional can be expressed in an alternative form \cite{Hameka:67}
\begin{equation}
E_2(L)= \langle \psi^{(a)} - \psi | H_0 - E_0 | \psi^{(a)} - \psi \rangle - \langle \psi | H_0 - E_0 | \psi \rangle
\end{equation}
which leads to
\begin{equation}
I_2(L) = \langle  \psi^{(b)} | H_0 - E_0|  \psi^{(b)}  \rangle \label{p2}
\end{equation}
As discussed in Section II.A, $\psi^{(b)} = O(r_{12}^{N+2})$.  Thus, Eq. (\ref{p2}) becomes
\begin{align}
  I_2(L) &=  \sum_{l=L}^{\infty}   \langle  \Phi_{N+2} R_{N+2,l} P_l | H_0 - E_0|   P_l R_{N+2,l}  \Phi_{N+2}\rangle \\
  &= \sum_{l=L}^{\infty}   \langle  \Phi_{N+2} R_{N+2,l} P_l | [H_0 - E_0, R_{N+2,l}] + R_{N+2,l} (H_0 - E_0) |  P_l  \Phi_{N+2} \rangle \\
  &= \sum_{l=L}^{\infty}   \langle  \Phi_{N+2} R_{N+2,l} P_l | -(N+2)(N+3) R_{N,l} + U_N + R_{N+2,l} (H_0 - E_0) |  P_l \Phi_{N+2}  \rangle
\end{align}
The increment is then
\begin{align}
  I_2(L+1) - I_2(L) &= \langle  \Phi_{N+2} R_{N+2,L} P_L | (N+2)(N+3) R_{N,L} - U_N - R_{N+2,L} (H_0 - E_0) |  P_L \Phi_{N+2}  \rangle
\end{align}
From the termwise analysis of the rates of convergence \cite{Wang:13}, we find that the leading term for the incremental rate of convergence is
\begin{equation}
(N+2) (N+3) \langle  \Phi_{N+2} R_{N+2,L} P_L | P_L R_{N,L}  \Phi_{N+2} \rangle
\end{equation}
\noindent It converges as $L^{-2N-6}$. A few expressions are reported in Table \ref{tab0}.

The rates of convergence must be the same from either Eqs. $(\ref{p1})$ or $(\ref{p2})$. Hence, we expect series of cancelations inside the expression (\ref{p1}). This effect will be verified in Section \ref{sec2d}.

\subsection{Formal solution of the first-order wavefunction and the relation to the assumptions of regularities of the wavefunction} \label{sec2c}

White and Brown obtained a  formal solution  \cite{White:67a,White:67c,White:67d} for the first-order wavefunction
\begin{align}
\psi &= \sum_{n=0}^{\infty} r_{12}^n \phi_n \label{wb0} \\
\phi_1 &= \frac{1}{2} \Phi \left[ 1 - \frac{1 }{3} s \eta^2 + \frac{1}{15} s^2 \eta^4 - \left(  \frac{1}{105} s^2 +  \frac{1}{105} s^3 \right) \eta^6 +\cdots \right] \label{wb1} \\
\phi_3 &= \frac{1}{9} \Phi \left[ \frac{1}{s} - \left(  \frac{7}{8s} + \frac{2}{5} \right)  \eta^2  +  \left( \frac{1}{16 s} +  \frac{251}{560} + \frac{3 s}{35 } \right) \eta^4 + \cdots \right] \label{wb2} \\
\phi_5 &=  \Phi \left[    \frac{13}{360 s^3} +  \frac{4}{225 s^2}  - \left( \frac{3}{40 s^3}  + \frac{68}{1575s^2} + \frac{4}{525s}  \right) \eta^2 + \cdots \right] \label{wb3} \\
\phi_7 &= \Phi \left[  \frac{37}{1680 s^5} + \frac{653}{58800 s^4} + \frac{8}{3675 s^3} +  \cdots  \right] \label{wb4}
\end{align}

\noindent where $s:=r_1 + r_2$ and $\eta:=(r_1 - r_2)/(r_1 + r_2)$.  We have corrected a typo in Ref. \cite{White:67a} and evaluated a few more odd-power terms. The second term in the square bracket of Eq. (\ref{wb1}) should be $-s \eta^2/3$, instead of $-s \eta^2$ in Ref. \cite{White:67a}. A comparison between the prefactors in the formal solution and variational values is presented in Table \ref{tab1}. The results from these two approaches are consistent.

The even-power $r_{12}$ terms contain logarithmic functions. Due to technical difficulties related to the boundary conditions, the explicit expressions have not been obtained \cite{White:67a,White:67c,White:67d}. Nevertheless, they have finite PWEs.

The odd-power $\phi_n$ in the formal solution may be regarded as $\Phi_n$ in our ansatz (\ref{Phi}) according to the uniqueness of the Taylor expansion.  The forms of odd-power terms lead to two issues:
\begin{itemize}
 \item[(i)] The presence of $- \frac{1}{6} s \eta^2 \Phi + \cdots$ term in Eq. (\ref{wb1}) seems to be in contradiction with the behavior (\ref{KM});
 \item[(ii)] The term with a negative power of $s$ may violate the assumptions of regularities in the first-order wavefunction \cite{Wang:13}.
\end{itemize}
We argue that point (i) is due to the ambiguity in Eq. (\ref{KM}). First, Schwartz's result is valid at $r_1=r_2 \neq 0$ and  the angular parameter $r_{12}$ is the only variable in the derivation \cite{Schwartz:62}. On the other hand, $\Phi$ in Eq. (\ref{KM}) contains radial information at $r_1 \neq r_2$. Second, KM derived Eq. (\ref{KM}) from the expansion of the first-order wavefunction near $r_{12}=0$ \cite{Kutzelnigg:92}. More specifically,
\begin{align}
\psi &= \sum_{l=0}^{\infty} \sum_{m=-l}^l \sum_{k=0}^{\infty} r_{12}^{l+k} Y_{lm} (\theta_{12},\phi_{12}) f_{lm}^{(k)}  \nonumber \\
&= \left[f_{00}^{(1)} r_{12} + O(r_{12}^2)  \right] Y_{00}  + \sum_{l=1,\cdots}^{\infty} \sum_{m=-l}^{l} \left[ f_{lm}^{(1)} r_{12} + O(r_{12}^2)  \right] r_{12}^l Y_{lm} \nonumber \\
&= \left[ \frac{1}{2}  \Phi_{00}^{(0)} r_{12}  + O(r_{12}^2)  \right] Y_{00}  + \sum_{l=1,\cdots}^{\infty} \sum_{m=-l}^{l} \left[ \frac{1}{2(l+1)} \Phi^{(0)}_{lm} r_{12} + O(r_{12}^2)  \right] r_{12}^l Y_{lm} \label{KM_e}
\end{align}
where we have followed the notations from Pack and Brown \cite{Pack:66}.  $f_{lm}^{(k)} $ is a function of other variables.  In the third line, the cusp conditions for $s$-wave ($l=0$) (\ref{cuspKM}) and other expansions ($l>0$) \cite{Pack:66,Kutzelnigg:92} were used. A derivation for the cusp conditions in the $1/Z$ expansion is supplemented in Appendix A.

The large-$l$ behavior (\ref{Sl}) and (\ref{KM}) correspond to $\frac{1}{2} \Phi_{00}^{(0)} Y_{00} r_{12}$ and $\frac{1}{2} \sum_{lm} \Phi_{lm}^{(0)} Y_{lm} r_{12} + O(r_{12}^2)$, respectively. Literally, we may denote both $\sum_{k \geq 3} f_{00}^{(k)} r_{12}^k Y_{00}$ and $\sum_{l\geq 2} \sum_{m=-l}^l \frac{1}{2(l+1)} \Phi^{(0)}_{lm} r_{12} r_{12}^l Y_{lm} $ as $O(r_{12}^3)$.  Nevertheless, a $r_{12}^l Y_{lm}$ term is conventionally factored out into non-$r_{12}$ functions. For example, $r_{12}^2 Y_{20} \propto 3 (z_2 - z_1)^2 + r_{12}^2$. Hence, the higher angular momentum terms resemble  $\left[ - \frac{1}{6} s \eta^2 + \cdots \right] r_{12} \Phi_1$ in Eq. (\ref{wb1}).

For point (ii), we notice the exact wavefunction is analytic away from the coalescence points \cite{Kato:57}. Near a two-particle-coalescence point it can be written as $\psi (\mathbf{x}) = \psi^{(1)}(\mathbf{x}) + | \mathbf{x} | \psi^{(2)} (\mathbf{x})$,
 where both $\psi^{(1)}(\mathbf{x})$ and $\psi^{(2)} (\mathbf{x})$ are analytic \cite{Fournais:09}. Hence, the term including a negative power of $s$, is expected to be arise from the three-particle-coalescence condition. It is not Taylor expandable at the origin. The previous work requires differentiability to $r_<$ for $\partial^{N+3} \psi / \partial r_{12}^{N+3}$, where $N$ is the highest power of geminal function \cite{Wang:13}.

If the singularities in the formal solution are the properties of the exact first-order wavefunction, it leads to either different rates of convergence than the inverse-power-law or weaker assumptions on the regularities of the wavefunction. We argue it is the second possibility and shall examine numerical results in Section III.

A Taylor expandable wavefunction implies a bounded remainder. We may formally proceed
\begin{align}
\frac{1}{s^n} =  \frac{1}{ \left( 2r_> \right)^n \left( 1 - \frac{ r_> - r_<}{2 r_>} \right)^n }  = \frac{1}{ (2 r_>)^n}  \left[ \sum_{m=0}^{\infty}  \left( \frac{r_>-r_<}{2 r_>}  \right)^m \right]^n \label{exp}
\end{align}

\noindent or regard the function with a negative power of $s$ as a basis.
In Hylleraas-type calculations, the integrals involving the basis fuction with a negative power of $s$ are finite \cite{Nakatsuji:07}. We may then expect the rate of convergence to remain
\begin{equation}
E_2(L) \rightarrow \sum_{n=n_{\min}}^{\infty} c_n (L+1/2)^{-n} \label{ser}
\end{equation}
\noindent If the absolute value of the prefactor $c_n$ increases slower than $(L+1/2)^n$,  Eq. (\ref{ser}) is converged. Otherwise, the series is asymptotic.

\subsection{Evaluations for the prefactors from the formal solution: computations for $L^{-8}$ term without the geminal function}  \label{sec2d}

The PWE without the geminal function is the simplest case to use the formal solution (\ref{wb0}) calculating the prefactors for the rates of convergence.  There are four schemes in the literature \cite{Schwartz:63,Byron:67,White:67b,Schmidt:83,Kutzelnigg:92,Jankowski:06}. Up to $L^{-6}$, the reported expressions are identical. We shall evaluate the prefactors at $L^{-8}$  from these approaches and compare them with numerical calculations. A few details are documented in Appendix B.

The first approach is an expansion in terms of $t/s$ at $\mathbf{r}_1 = \mathbf{r}_2$ \cite{Schwartz:63,White:67b}.  Formulae up to $L^{-6}$ have been reported \cite{Schwartz:63,Byron:67,White:67b}. It is straightforward to obtain  expression up to $L^{-8}$
\begin{equation}
E_2(L) - E_2(L-1) = - \frac{45}{256} (L+1/2)^{-4}  + \frac{225}{1024} (L+1/2)^{-6} -\frac{15345}{16384}(L+1/2)^{-8} + \cdots \label{s8}
\end{equation}

The second method \cite{Kutzelnigg:92} is to regard the wavefunction (\ref{KM}) as the asymptotic behavior
\begin{equation}
\chi_l \rightarrow \frac{1}{2} R_{1l}  \Phi +  \frac{1}{9s}R_{3l} \Phi + \cdots \label{sll}
\end{equation}
where we have adopted the formal solution (\ref{wb2}) as $O(r_{12}^3)$. Along this line, the result is
\begin{equation}
E_2(L) - E_2(L-1) = - \frac{45}{256} (L+1/2)^{-4}  + \frac{225}{1024} (L+1/2)^{-6} -\frac{1785}{16384}(L+1/2)^{-8} + \cdots \label{km8}
\end{equation}

If the asymptotic behavior of the first-order wavefunction, which determines up to the $L^{-8}$ partial-wave increment, is the formal solution (\ref{wb0})
\begin{equation}
\chi_l \rightarrow \frac{1}{2} \left[ 1 - \frac{1}{3} s \eta^2 \right] R_{1l}  \Phi  +  \frac{1}{9s} R_{3l}\Phi + \cdots \label{sll-2}
\end{equation}

\noindent $- \frac{1 }{6} s \eta^2 r_{12} \Phi$ will provide a $-735/1024(L+1/2)^{-8}$ correction.
The total energy increment from the asymptotic behavior (\ref{sll-2}) is then identical to the first approach. From the above calculations, any term proportional to $ s \eta^2 \Phi$ in $\chi$ will not affect up to $L^{-6}$ rates of convergence.

The third method \cite{Schmidt:83} is to regard the wavefunction (\ref{Sl}) as the large-$l$ limit. The energy increment is determined by transforming the expression of $E_2(L)$ into a James-Coolidge type integral. We are unable to understand this approach in detail. We list their result and will compare it with numerical results.
\begin{equation}
E_2(L) - E_2(L-1) = - \frac{45}{256} (L+1/2)^{-4}  + \frac{225}{1024} (L+1/2)^{-6}  -\frac{8235}{16384}(L+1/2)^{-8} + \cdots \label{sh8}
\end{equation}

The fourth method \cite{Jankowski:06} is used in evaluating the rates of convergence in the second-order M{\o}ller-Plesset perturbation. The idea can also be applied to the $1/Z$ expansion. Instead of the Hylleraas functional, one starts from the expression of the second-order energy
\begin{equation}
E_2 = \langle \psi | H' | \Phi \rangle
\end{equation}
and use the large-$l$ behavior of $\psi$ to derive $E_2(L)$. As mentioned by the authors, the somewhat complicated $D(L)$ term in KM's approach \cite{Kutzelnigg:92}  does not appear. However, this scheme requires higher-order terms in $\chi_l$ to determine $E_2(L)$. $\chi_l \rightarrow \frac{1}{2} R_{1l} \Phi$ will fix the expression of $E_2(L)$ up to $L^{-4}$:
\begin{equation}
E_2(L) - E_2(L-1) = - \frac{45}{256} (L+1/2)^{-4}  - \frac{225}{2048} (L+1/2)^{-6} -\frac{1575}{8192}(L+1/2)^{-8} + \cdots \label{km8v1}
\end{equation}
\noindent while the Hylleraas functional can fix $E_2(L)$ to $L^{-6}$ \cite{Kutzelnigg:92}. This may be due to the variational method can approximate the energy in the error of $O(\varepsilon^2)$ for the wavefunction with $O(\varepsilon)$ uncertainty \cite{Helgaker:2000}. Including $-\frac{1}{6} s\eta^2 R_{1l} \Phi$ and $\frac{1}{9s}  R_{3l} \Phi$ as Eq. (\ref{sll-2}) will convergence $E_2(L)$ up to $L^{-6}$
\begin{equation}
E_2(L) - E_2(L-1) = - \frac{45}{256} (L+1/2)^{-4}  + \frac{225}{1024} (L+1/2)^{-6} + \frac{2205}{2048}(L+1/2)^{-8} + \cdots \label{km8v2}
\end{equation}
Introducing additional terms in $\chi_l$ as
\begin{equation}
\chi_l \rightarrow \frac{1}{2} \left[ 1 - \frac{1 }{3} s \eta^2  + \frac{1}{15} s^2 \eta^4 \right] R_{1l}  \Phi  +  \frac{1}{9} \left[ \frac{1}{s} - \left(\frac{7}{8s} + \frac{2}{5} \right) \eta^2 \right] R_{3l} \Phi + \left[     \frac{13}{360s^3} + \frac{4}{225 s^2} \right] R_{5l} \Phi
\end{equation}
$E_2(L)$ becomes identical to Eq. (\ref{s8}) up to $L^{-8}$.

A comparison between formulae (\ref{s8}), (\ref{km8}), and (\ref{sh8}) and numerical results \cite{Wang:13} is given in Table \ref{tab2}. As we see, Eq. (\ref{s8}) provides the best agreement with the numerical calculations. It is thus supported that large-$l$ behavior should be (\ref{sll-2}).

\subsection{Evaluations of the prefactors from the formal solution: computations for the prefactors with the geminal functions} \label{sec2e}
Based on the formal solution (\ref{wb0}), if we adopt the ansatz (\ref{ansatz1}), the large-$l$ behavior of $\chi$ is
\begin{equation}
\chi_l \rightarrow  \frac{\Phi }{9s} R_{3l} - \frac{\Phi}{6} s \eta^2   R_{1l}
\end{equation}

The first and second terms here converge as $-735/1024(L+1/2)^{-8} \approx -0.72 (L+1/2)^{-8}$ and $-3675/16384(L+1/2)^{-8} \approx -0.22 (L+1/2)^{-8}$, respectively. The total contribution, $-15345/16384(L+1/2)^{-8} \approx -0.94(L+1/2)^{-8}$,  agrees with the numerical fittings $-0.88 (L+1/2)^{-8}$ \cite{kutzelnigg1985r} and $-0.88(L+0.56)^{-8}$ \cite{Wang:13}.

We can adopt a modified ansatz to include  $-\frac{s \eta^2}{6} r_{12} \Phi$ and other linear $r_{12}$ terms in Eq. (\ref{wb1})
\begin{align}
\psi^{(a)}  &=   r_{12} \Phi_1  +  \sum_{l=0}^L \psi_l(r_1,r_2) P_l (\cos \theta) \label{a1p} \\
\Phi_1 &= \phi_1 \,\,\,\,\,\,\,\,\,\,
\end{align}

As presented in Table \ref{tab2}, from the first expression of $I_2(L)$, Eq. (\ref{p1}), there are three terms contribute to the $L^{-8}$ rates of convergence. From the second expression of $I_2(L)$, Eq. (\ref{p2}), all contributions are merged into a single term. Previously, the explanation of the $r_{12}$ function leading to the $L^{-8}$ rate of convergence is to let the regular operator $U_1$ replace the singular operator $r_{12}^{-1}$ \cite{kutzelnigg1985r}. As we see in this and the following examples, it is increasingly simpler to think from (\ref{p2}) for the leading order of the rates of convergence. Namely, the Legendre polynomial is not efficient to span the singular function at the origin. Introducing odd-power geminal functions will let the Legendre polynomials span more regular functions, i.e. higher power of $r_{12}$ functions.

The term $ - \frac{s \eta^2}{3} r_{12} \Phi$ was missing in our previous study \cite{Wang:13}. The large-$l$ PWE convergence will be dominated by $s \eta^2 R_{1l} \Phi$, which contributes to $L^{-8}$. Hence, the previous fitted $L^{-10}$ and $L^{-12}$ lines are mixtures of $L^{-8}$, higher inverse power rates of convergence, and relaxations of amplitudes in variational calculations.

We consider an extended ansatz
\begin{align}
\psi^{(a)}  &=   r_{12} \Phi_1  +  r_{12}^3 \Phi_3 + \sum_{l=0}^L \psi_l(r_1,r_2) P_l (\cos \theta)  \label{a4} \\
\Phi_1 &=  \phi_1, \,\,\,\,\,\,\,\,\,\, \Phi_3 = \phi_3
\end{align}
\noindent to include the missing terms at the $r_{12}$ level. The evaluated PWE prefactors are presented in Table \ref{tab4}.
As expected, all contributions at $L^{-10}$ vanishes, leaving identical $L^{-12}$ expressions given by Eq. (\ref{p2}). The case with $r_{12}^5$
\begin{align}
\psi^{(a)}  &=   r_{12} \Phi_1  +  r_{12}^3 \Phi_3  +  r_{12}^5 \Phi_5 +  \sum_{l=0}^L \psi_l(r_1,r_2) P_l (\cos \theta)  \label{a5} \\
\Phi_1 &=  \phi_1, \,\,\,\,\,\,\,\,\,\, \Phi_3 = \phi_3, \,\,\,\,\,\,\,\,\,\, \Phi_5 = \phi_5
\end{align}
is similar. In the supplementary material, a Maple code \cite{MAPLE:11} was provided for the evaluation of prefactors in Tables \ref{tab3} - \ref{tab6}.

\begin{table}[htbp]
  \centering
  \caption{Leading expressions of the rates of convergence for both second-order $1/Z$ and non-perturbed energies of the ground state helium atom. }
    \begin{tabular}{ll}     \hline
$N$             & \hspace{-6cm} Expression \\ \hline
0               &  \hspace{-6cm} $  -\frac{24 \pi^2}{(l+1/2)^4} \int_0^{\infty} \left| \frac{ \partial \psi (r,r,0)}{\partial r_{12}} \right|^2 r^5 dr^a  $ \\
1               & \hspace{-6cm} $  -\frac{2520 \pi^2}{(l+1/2)^8} \int_0^{\infty} \left| \frac{ \partial^3 \psi (r,r,0)}{\partial r_{12}^3} \right|^2 r^9 dr $ \\
3               & \hspace{-6cm} $  -\frac{1871100 \pi^2}{(l+1/2)^{12}} \int_0^{\infty} \left| \frac{ \partial^5 \psi (r,r,0)}{\partial r_{12}^5 } \right|^2 r^{13} dr $ \\
5               & \hspace{-6cm}$  -\frac{5108103000 \pi^2}{(l+1/2)^{16}} \int_0^{\infty} \left| \frac{ \partial^7 \psi (r,r,0)}{\partial r_{12}^7} \right|^2 r^{17} dr $ \\ \hline
$^a$  By the cusp condition, we obtain  $-\frac{6 \pi^2}{(l+1/2)^4} \int_0^{\infty} \left| \psi (r,r,0) \right|^2 r^5 dr $
    \end{tabular}%
  \label{tab0}%
\end{table}%

\begin{table}[htbp]
  \centering
  \caption{Comparison between formal and variational optimized solutions for the first order wavefunction of the ground state helium atom. In the variational calculations, we have used basis B and F for the odd-power and even-power of $r_{12}$ functions, respectively \cite{Schwartz:06a}. Parameters $\omega=20$ and $\alpha=1$ are adopted in the trial wavefunction. }
    \begin{tabular}{rrrrrr}     \hline
    Term   &   \multicolumn{2}{c}{ Coefficient} &  Term   &   \multicolumn{2}{c}{ Coefficient}\\
      & Formal solution & Variational result       &   & Formal solution & Variational result  \\ \hline
    $r_{12} \Phi$    &  $0.500$                             &  0.500 & $r_{12}^5 s^{-3} \Phi$ &   $0.0361 $ &  0.0361 \\
    $r_{12} s \eta^2 \Phi$       &       $-0.167$           &  -0.167 & $r_{12}^5 s^{-2} \Phi$  & $0.0178$ & 0.0178  \\
    $r_{12} s^2 \eta^4 \Phi$       &       $0.0333$         &  0.0336 & $r_{12}^5 s^{-3} \eta^2 \Phi$   &  $-0.0750$   &   -0.0763  \\
    $r_{12} s^2 \eta^6 \Phi$       &       $-0.00476$       & -0.00468  & $r_{12}^5 s^{-2} \eta^2 \Phi$  &  $-0.0432$ & -0.0426   \\
    $r_{12} s^3 \eta^6 \Phi$       &       $-0.00476$       & -0.00471  & $r_{12}^5 s^{-1} \eta^2 \Phi$   &  $-0.00762$  & -0.00771 \\
    $r_{12}^3 s^{-1} \Phi$       &       $0.111$            & 0.111  & $r_{12}^{7} s^{-5}  \Phi$ & $0.0220$  & 0.0226\\
    $r_{12}^3 s^{-1} \eta^2 \Phi$       &       $-0.0972$   & -0.0973   & $r_{12}^{7} s^{-4} \Phi$  & $0.0110 $  & 0.0108  \\
    $r_{12}^3        \eta^2 \Phi$       &       $-0.0444$   &  -0.0444 &  $r_{12}^{7} s^{-3} \Phi$ &  $0.00218$ & 0.00222 \\
    $r_{12}^3  s^{-1} \eta^4 \Phi$       &       $0.00694$  & 0.00816  & & &  \\
    $r_{12}^3         \eta^4 \Phi$       &      $0.0498$&  0.0491    \\
    $r_{12}^3   s      \eta^4 \Phi$       &      $0.00952$  &    0.00972  \\
     \hline
    \end{tabular}%
  \label{tab1}%
\end{table}%

\begin{table}[htbp]
  \centering
  \caption{Comparison between theoretical and numerical rates of convergence for the partial-wave increments of the second-order $1/Z$ energy. The boldface digits in the numerical results indicate the values are expected to be converged.  }
    \begin{tabular}{lrrr}  \hline
    $L$    & 10   & 100  & 1000  \\  \hline
    Eq. (\ref{s8})       & -1.43031 99170  $\times 10^{-5}$     &     -1.72287 81776 $\times 10^{-9}$  &     -1.75429 90744 63447  $\times 10^{-13}$   \\
    Eq.  (\ref{km8})    &  -1.42983 40997  $\times 10^{-5}$       &   -1.72287 81086 $\times 10^{-9}$     &  -1.75429 90744 56297  $\times 10^{-13}$ \\
    Eq.  (\ref{sh8})    &  -1.43010 05556   $\times 10^{-5}$        &  -1.72287 81464  $\times 10^{-9}$   &   -1.75429 90744 60219  $\times 10^{-13}$   \\
    Numer. \cite{Wang:13}   & \textbf{-1.43029 20705}$\times 10^{-5}$   &  \textbf{-1.72287 81775}  $\times 10^{-9}$     &   \textbf{-1.75429 907}44 63457  $\times 10^{-13}$    \\  \hline
    \end{tabular}%
  \label{tab2}%
\end{table}%

\begin{table}[htbp]
  \centering
  \caption{Leading expressions for the rates of convergence based on ansatz (\ref{ansatz1}), $\Phi_1 = \Phi$.  }
    \begin{tabular}{lr}     \hline
Term      & Rate of convergence \\ \hline
  $   \langle- \frac{1}{6} s \eta^2 \Phi R_{1l} P_l | U_1 | \Phi_1 \rangle $    & $-\frac{3675}{16384} (L+1/2)^{-8}  $   \\
     $   \langle \frac{1}{9s} \Phi R_{3l} P_l | U_1 | \Phi_1 \rangle $    & $ -\frac{735}{1024}(L+1/2)^{-8} $  \\
& \\
Sum & $-\frac{15345}{16384} (L+1/2)^{-8} $  \\
     \hline
    \end{tabular}%
  \label{tab3}%
\end{table}%

\begin{table}[htbp]
  \centering
  \caption{Leading expressions for the rates of convergence based on ansatz (\ref{a1p}), $\Phi_1=\phi_1,\chi_l \rightarrow \phi_3 R_{3l} $.  }
    \begin{tabular}{lr}     \hline
Term      & Rate of convergence \\ \hline
  $   \langle \chi  | U_1 | \Phi_1 \rangle^{a} $    & $-\frac{2205}{4096} (L+1/2)^{-8 } $   \\
   $   \langle \chi  | r_{12}^{-1} | - 2 \Phi_1 + \Phi \rangle $    & $\frac{735}{8192} (L+1/2)^{-8} $  \\
   $  \langle  \chi | r_{12} (H_0 - E_0) | \Phi_1 \rangle $  &  $ -\frac{1225}{8192}(L+1/2)^{-8}  $ \\
Sum & $-\frac{1225}{2048} (L+1/2)^{-8} $  \\  \\
$  12    \langle  \phi_3 R_{3l} | R_{1l}  \phi_3 \rangle $   & $ -\frac{1225}{2048}(L+1/2)^{-8}  $  \\ \hline
   & \hspace{-8.3cm} $^a$ Both $\eta^0$ and $\eta^2$ in $\Phi_1$ contribute to the leading order of the rate of convergence.  \\
    \end{tabular}%
  \label{tab4}%
\end{table}%

\begin{table}[htbp]
  \centering
  \caption{Leading expressions for the rates of convergence based on ansatz  (\ref{a4}), $\Phi_1=\phi_1, \Phi_3=\phi_3, \chi_l \rightarrow \phi_5 R_{5l} $.}
    \begin{tabular}{ll}     \hline
Term      & Rate of convergence \\ \hline
    $ \langle \chi | U_1 | \Phi_1 \rangle $    &  $\frac{936243}{262144} (L+1/2)^{-10}    +   \frac{18904545}{524288} (L+1/2)^{-12}$   \\
    $  -12  \langle \chi | r_{12} | \Phi_3 \rangle $  &  $ -\frac{255339}{65536}(L+1/2)^{-10}  -  \frac{111296097}{2621440}  (L+1/2)^{-12} $ \\
    $ \langle \chi  | r_{12}^{-1} | -2   \Phi_1 + \Phi \rangle $  &  $ -\frac{85113}{131072}(L+1/2)^{-10} -  \frac{1276209}{524288}  (L+1/2)^{-12}  $  \\
    $   \langle \chi | r_{12} (H_0 - E_0) | \Phi_1  \rangle $  &  $ \frac{255339}{262144}(L+1/2)^{-10}  + \frac{11732787}{1310720}(L+1/2)^{-12}   $  \\
 $ \langle \chi  | U_3 | \Phi_3 \rangle $    &  $-\frac{37593171}{1310720} (L+1/2)^{-12} $   \\
   $   \langle  \chi | r_{12}^3 (H_0 - E_0) | \Phi_3  \rangle $  &  $ \frac{27112239}{20971520}  (L+1/2)^{-12} $  \\
Sum &  $-\frac{571889241}{20971520} (L+1/2)^{-12}$ \\
& \\
$  30    \langle  \phi_5 R_{5l} | R_{3l}\phi_5 \rangle $   & $ -\frac{571889241}{20971520} (L+1/2)^{-12}  $  \\
     \hline
    \end{tabular}%
  \label{tab5}%
\end{table}%

\begin{table}[htbp]
  \centering
  \caption{Leading expressions for the rates of convergence based on ansatz  (\ref{a5}), $\Phi_1=\phi_1, \Phi_3=\phi_3,\Phi_5=\phi_5, \chi_l \rightarrow \phi_7 R_{7l} $.}
    \begin{tabular}{ll}     \hline
Term      & Rate of convergence \\ \hline
    $ \langle  \chi  | U_1 | \Phi_1 \rangle $    &  $ \,\,\,\,\, - {\frac {23980671}{524288}} (L+1/2)^{-12} - {\frac {151194341727}{134217728}}  (L+1/2)^{-14} -{\frac {147653411808825}{
5905580032}} (L+1/2)^{-16} $  \\
    $  -12  \langle \chi | r_{12} | \Phi_3 \rangle $  & ${\frac {12912669}{262144}} (L+1/2)^{-12} +  {\frac {21304170459}{16777216 }} (L+1/2)^{-14}         + {\frac {29642661896679}{1073741824}} (L+1/2)^{-16} $ \\
    $  \langle \chi | r_{12}^{-1} | -2  \Phi_1 + \Phi \rangle $  &    $ {\frac {9223335}{1048576}} (L+1/2)^{-12}   +  {\frac {9986230683}{67108864} (L+1/2)^{-14}} +{\frac {23836173850125}{5905580032}}  (L+1/2)^{-16} $ \\
    $   \langle \chi | r_{12} (H_0 - E_0) | \Phi_1  \rangle $  &  $ -{\frac {12912669}{1048576}}  (L+1/2)^{-12}  -  {\frac {39530417355}{134217728}}  (L+1/2)^{-14}         -{\frac {3528771151581}{
536870912}}  (L+1/2)^{-16}$ \\
 $ \langle  \chi | U_3 | \Phi_3 \rangle $   &  $\frac{12928523751}{33554432}  (L+1/2)^{-14}   + \frac{11125016454123}{1073741824}  (L+1/2)^{-16}   $ \\
 $   \langle \chi | r_{12}^3 (H_0 - E_0) | \Phi_3  \rangle $   &  $- \frac{838111131}{41943040}  (L+1/2)^{-14}   - \frac{1447496015823}{2147483648}  (L+1/2)^{-16}$  \\
 $ \langle  \chi | U_5 | \Phi_5 \rangle $   &  $  - \frac{411530302755}{67108864}  (L+1/2)^{-16}   $ \\
    $  -30  \langle  \chi | r_{12}^3 | \Phi_5 \rangle $  & $ -\frac{15222876669}{41943040} (L+1/2)^{-14} -  \frac{21033724282857}{2147483648 } (L+1/2)^{-16}         $ \\
    $    \langle \chi | r_{12}^5 (H_0 - E_0) | \Phi_5 \rangle $  & $   \frac{309328265961}{268435456 } (L+1/2)^{-16}         $ \\
Sum & $-\frac{673733870667}{134217728 }  (L+1/2)^{-16} $ \\
&  \\
$  56    \langle   \phi_7 R_{7l} |   R_{5l} \phi_7 \rangle $     & $ -\frac{673733870667}{134217728 }  (L+1/2)^{-16}   $  \\
     \hline
    \end{tabular}%
  \label{tab6}%
\end{table}%

\section{Numerical results for the PWE in the second-order energies in the $1/Z$ expansions} \label{sec3}

In Table \ref{tab70}, we present the numerical results for the PWE
\begin{equation}
\psi = \sum_{n=1,3,\cdots}^{N} r_{12}^n \Phi_n + \sum_{l=0,1,\cdots}^L  P_l(\cos \theta_{12}) \chi_l \label{span1}
\end{equation}

\noindent from spanning the reference wavefunction. The reference wavefunction is optimized from basis F in Schwartz's article \cite{Schwartz:06a}
\begin{align}
\psi &= \sum_{i=1}^{M_n} c_i \phi_i \label{SS} \\
\phi &= s^{\lambda} (t/s)^{\mu} (u/s)^{\nu} (\ln s)^{\zeta} e^{-\alpha s} \label{SS2}
\end{align}

In Eq. (\ref{span1}), $\Phi_n$ is taken as the relevant odd-power terms of $r_{12}$ in the reference wavefunction (\ref{SS}).  $\chi_l$ is converted from other $r_{12}$ terms by the Sack expansion \cite{Sack:64,Cohl:13}. The obtained quantities correspond to $E_2(L)$ and $\chi_l$.

\begin{figure}
\begin{center}
\includegraphics[width=12cm,angle=-90]{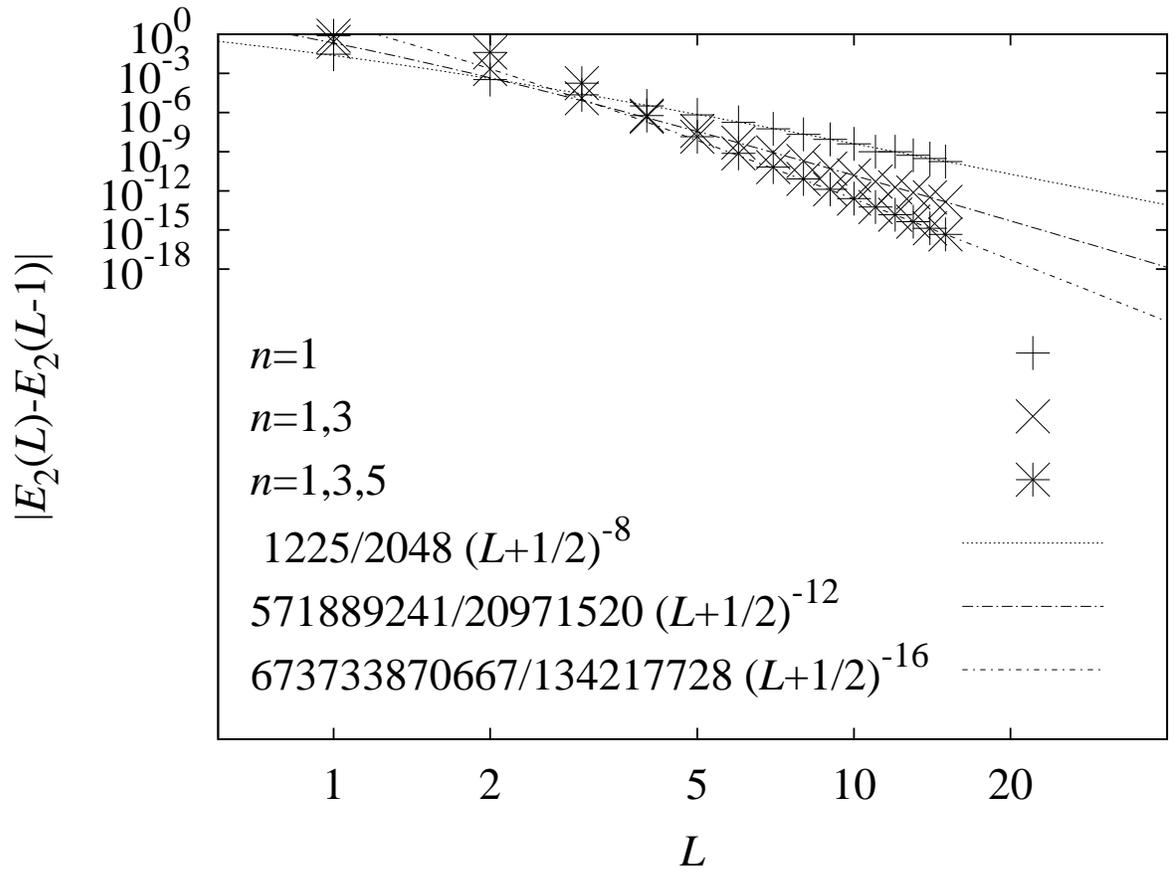}
\caption{$|E_2(L) - E_2(L-1)|$ via $L$ plot for the second-order $1/Z$ expansion energy. The lines are obtained from the equations in Tables \ref{tab4} - \ref{tab6}. The numerical data points are obtained from Table \ref{tab70}. Since the partial-wave increment is taken in absolute value, the prefactors of the rates of convergence are positive. Atomic units are used in the figure. }
\label{fig_1z0}
\end{center}
\end{figure}

In Figure \ref{fig_1z0}, the numerical data points agree well with the theoretical rates of convergence in dashed lines. In addition, we compared the rates of convergence from variational optimized increments in the supplementary material. The convergence is similar to the expanded values.

\begin{table}[htbp]
\renewcommand{\tabcolsep}{0.3cm}
  \centering
  \caption{Numerical results of the PWE increments for the second-order-$1/Z$ expansion energies of the ground state of the helium atom. Eqs. (\ref{SS}) and (\ref{SS2}) are used as the wavefunction at $\omega=20$ in the Hylleraas functional \cite{Schwartz:06a}. The atomic units are adopted.}
    \begin{tabular}{llll}
    \hline
    $L$     &       \multicolumn{3}{c}{$ |E_2(L) - E_2(L-1)|$}                          \\
            & $r_{12} \Phi_1 +  \chi$   & $r_{12} \Phi_1 + r_{12}^3 \Phi_3 + \chi$   & $r_{12} \Phi_1 + r_{12}^3 \Phi_3 + r_{12}^5 \Phi_5+ \chi$   \\
    \hline
    0$^a$     &1.283$\times10^{-1}$  & 1.031$\times10^{-1 b}$  & 5.722$\times10^{-1b }$  \\
    1     &2.904$\times10^{-2}$  & 2.586$\times10^{-1}$  & 6.907$\times10^{-1}$\\
    2     &3.300$\times10^{-4}$  & 2.141$\times10^{-3}$  & 3.900$\times10^{-2}$\\
    3     &2.284$\times10^{-5}$  & 1.222$\times10^{-5}$  & 1.718$\times10^{-4}$     \\
    4     &3.173$\times10^{-6}$  & 4.654$\times10^{-7}$  & 5.753$\times10^{-7}$  \\
    5     &6.553$\times10^{-7}$  & 3.835$\times10^{-8}$  & 1.377$\times10^{-8}$       \\
    6     &1.756$\times10^{-7}$  & 4.977$\times10^{-9}$  & 7.490$\times10^{-10}$      \\
    7     &5.669$\times10^{-8}$  & 8.780$\times10^{-10}$ & 6.669$\times10^{-11}$            \\
    8     &2.104$\times10^{-8}$  & 1.937$\times10^{-10}$ & 8.331$\times10^{-12}$     \\
    9     &8.709$\times10^{-9}$  & 5.075$\times10^{-11}$ & 1.337$\times10^{-12}$ \\
    10    &3.933$\times10^{-9}$  & 1.523$\times10^{-11}$ & 2.606$\times10^{-13}$\\
    11    &1.908$\times10^{-9}$  & 5.103$\times10^{-12}$ & 5.934$\times10^{-14}$\\
    12    &9.827$\times10^{-10}$ & 1.874$\times10^{-12}$ & 1.535$\times10^{-14}$  \\
    13    &5.325$\times10^{-10}$ & 7.439$\times10^{-13}$ & 4.422$\times10^{-15}$  \\
    14    &3.013$\times10^{-10}$ & 3.155$\times10^{-13}$ & 1.395$\times10^{-15}$   \\
    15    &1.771$\times10^{-10}$ & 1.417$\times10^{-13}$ & 4.759$\times10^{-16}$   \\
          &                      &                            &   \\
  $\Delta E^c$ &    3.123 $\times10^{-10}$  &      1.380 $\times10^{-13}$          &    2.911$\times10^{-16}$ \\
    \hline
&&& \hspace{-8.3cm} $^a$ The energy of $|E_2(L)|$, $L$=0. \\
&&& \hspace{-8.3cm} $^b$ The energy is positive. \\
&&& \hspace{-8.3cm} $^c$ The total energy at the largest $L$ subtracts the reference value, -0.15766 64294 69150 94105 66 a.u. \cite{Wang:13}  \\
\end{tabular}%
\label{tab70}
\end{table}%

\section{Rates of convergence in the Ritz's variational approach} \label{sec4}

We start from a recapitulation of the previous work \cite{Wang:13}.
\begin{align}
\Delta E(L) := E(L) - E  &=  \frac{ \langle \psi - \psi^{(b)} | H - E | \psi - \psi^{(b)} \rangle }{ \langle \psi  -  \psi^{(b)}  | \psi  -  \psi^{(b)} \rangle} \label{vp}
\end{align}
\noindent where $\psi$ and $E$ represent the exact wavefunction and energy, respectively. The definitions of $\psi^{(a)}$ and $\psi^{(b)}$ are similar to the $1/Z$ expansion
\begin{align}
\psi &= \psi^{(a)} + \psi^{(b)} \\
 \psi^{(a)}  &:=   \psi_g  +  \chi  \,\,\,\,\,\, \\
\psi_g  &:=  \sum_{n=1,3,5,\cdots}^{N} r_{12}^n \Phi_n (r_1,r_2), \,\,\,\,\,\, \chi := \sum_{l=0}^{L} P_l (\cos \theta_{12}) \chi_l (r_1,r_2)  \\
\Phi_n &:= \frac{1}{n!} \left[ \sum_{m} \frac{ (-|r_1 - r_2|)^m }{m! \,\,} \left. \frac{ \partial^{n+m} \psi }{\partial r_{12}^{n+m} }\right|_{\substack{ r_{12}= |r_1 - r_2|}} \right] \\
\chi_l &:= \frac{2l+1}{2} \int_0^{\pi} \big[ \, \psi (r_1, r_2, r_{12}) - \sum_{n=1,3,5,\cdots}^{N} r_{12}^n \Phi_n(r_1, r_2) \, \big] P_l (\cos \theta_{12}) \sin \theta_{12} d \theta_{12} \\
\psi^{(b)} &= \sum_{l={L+1}}^{\infty} P_l (\cos \theta_{12}) \chi_l (r_1,r_2)
\end{align}

\noindent In Eq. (\ref{vp}) the exact wavefunction $\psi$ is normalized.

The variation in the denominator in Eq. (\ref{vp}) contributes to the rates of convergence beyond the leading order \cite{Wang:13}. Since the rates of convergence from the electron-electron repulsion is faster than the kinetic term \cite{Hill:85,Wang:13}, the PWE is determined by
\begin{align}
I(L)= \langle \psi^{(b)} | H_0- E | \psi^{(b)}  \rangle \label{vpf}
\end{align}

\noindent here $H_0:=H-1/r_{12}= -\frac{1}{2} \nabla_1^2  -\frac{1}{2} \nabla_2^2 - \frac{Z}{r_1} - \frac{Z}{r_2} $.

In the previous work \cite{Wang:13}, we performed further derivations to convert Eq. (\ref{vpf}). The "$L^{-N-7}$" rates of convergence were obtained.  By a similar derivation with Eq. (\ref{p2}) in the $1/Z$ expansion, Eq. (\ref{vpf}) converges as $L^{-2N-6}$. The leading rates of convergence have the same functional forms as in the $1/Z$ expansion.
\section{Numerical results for the PWE in Ritz's variations} \label{sec5}

\begin{figure}
\begin{center}
\includegraphics[width=12cm,angle=-90]{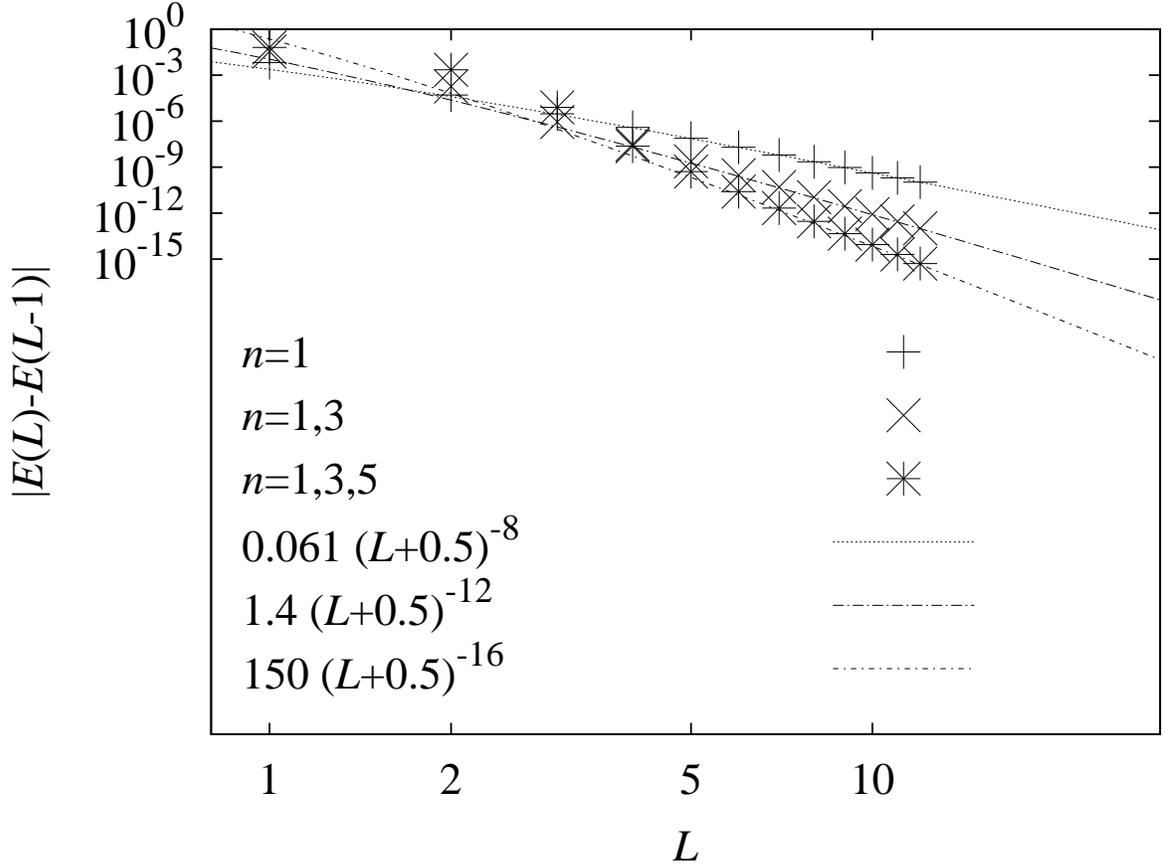}
\caption{$|E(L) - E(L-1)|$ via $L$ plot for the PWE from the reference wavefunction. The data of the PWE is plotted from Table \ref{ci_f12}. The prefactors of lines are obtained according to the formulae in Table \ref{tab0} and with $\omega=14$ of Basis F \cite{Schwartz:06a}. Atomic units are used in the figure. }
\label{fig_vp0}
\end{center}
\end{figure}

In Figure \ref{fig_vp0} we present comparisons between the theoretical and variational calculations. The prefactors are determined by basis F \cite{Schwartz:06a} with $\omega=14$. The expanded PWEs in Figures \ref{fig_vp0} agree well with the theoretical rates of convergence.

\begin{table}[htbp]
  \centering
  \caption{Rates of convergence of the PWE increments for the Ritz variational energy of the ground state helium atom by expanding the reference wavefunction. Basis F at $\omega=14$ is used for the reference wavefunction. The atomic units are adopted.}
    \begin{tabular}{llll}
    \hline
    $L$     &       \multicolumn{3}{c}{$ |E(L) - E(L-1)|$}                          \\
            & $\Phi_0 + r_{12}\Phi_1 + \chi$   & $\Phi_0 + r_{12} \Phi_1 + r_{12}^3 \Phi_3 + \chi$   & $\Phi_0 + r_{12} \Phi_1 + r_{12}^3 \Phi_3 + r_{12}^5 \Phi_5+ \chi$   \\
    \hline
    0$^a$  & 2.897   &   2.863     &  2.838   \\
    1     &6.624$\times10^{-3}$  & 3.456$\times10^{-2}$ & 6.373$\times10^{-2}$\\
    2     &5.044$\times10^{-5}$  & 1.837$\times10^{-4}$ & 2.239$\times10^{-3}$\\
    3     &3.002$\times10^{-6}$  & 8.729$\times10^{-7}$ & 8.003$\times10^{-6}$      \\
    4     &3.863$\times10^{-7}$  & 3.024$\times10^{-8}$ & 2.392$\times10^{-8}$ \\
    5     &7.625$\times10^{-8}$  & 2.354$\times10^{-9}$ & 5.340$\times10^{-10}$        \\
    6     &1.985$\times10^{-8}$  & 2.943$\times10^{-10}$& 2.771$\times10^{-11}$\\
    7     &6.283$\times10^{-9}$  & 5.060$\times10^{-11}$& 2.386$\times10^{-12}$\\
    8     &2.299$\times10^{-9}$  & 1.096$\times10^{-11}$& 2.907$\times10^{-13}$  \\
    9     &9.416$\times10^{-10}$ & 2.832$\times10^{-12}$& 4.574$\times10^{-14}$   \\
   10     &4.218$\times10^{-10}$ & 8.406$\times10^{-13}$& 8.774$\times10^{-15}$ \\
   11     &2.033$\times10^{-10}$ & 2.794$\times10^{-13}$& 1.972$\times10^{-15}$ \\
          &                      &                      &  \\
  $\Delta E^b$ & $2.432\times10^{-10}$      & 1.743 $\times10^{-13}$           &  8.679$\times10^{-15}$ \\
    \hline
&&& \hspace{-8cm} $^a$ The energy of $|E(L)|$, $L$=0. \\
&&& \hspace{-8cm} $^b$ Total energy at the largest $L$ subtracts the reference value - 2.903 724 377 034 119 598 311 \cite{Schwartz:06b,kurokawa2008solving}.
    \end{tabular}
\label{ci_f12}
\end{table}

\section{Summary}

We report the progress on evaluating the prefactors for the rates of convergence in the PWE with the presence of the odd-power geminal functions.
With the formal solution from White and Brown \cite{White:67a}, analytic expressions of the prefactors for the second-order $1/Z$ expansion energy in the PWE were obtained.

Series of cancelations of the inverse power law at $L^{-N-7}$ were found. The correct rates of convergence were $L^{-2N-6}$ for both the second-order energy of the $1/Z$ expansion and the energy of the Ritz's variational approach.

For more general systems, such as a lithium atom, the linked term $r_{12} r_{13}$ will bring additional features in the convergence of the PWE. The Pauli exclusive principle restricts the three electrons  in different configurations. The Fermi hole between the electrons in parallel spins will repel each other. This correlation is expected to be less significant than the power series of $r_{12}$.

\appendix

\section{The cusp conditions for the first-order wavefunction}

The cusp conditions for the first-order $1/Z$ wavefunction have been given in KM's paper \cite{Kutzelnigg:92}. To support our discussion in Section II.C, we supplement a few details of the derivation.

Follow Pack and Brown's approach \cite{Pack:66}, we start from the first-order equation $(H_0 - E_0) \psi = -(H' - E_1) \Phi $ near $r_{12}=0$,
\begin{align}
\left[ - \frac{1}{2 \mu} \left(  \frac{1}{r^2}  \frac{ \partial }{ \partial r} r^2   \frac{ \partial}{\partial r}  - \frac{L^2}{r^2} \right) + O(r^0) \right] \psi &= \left[ - \frac{1}{r} + O(r^0)  \right] \Phi \label{A1}
\end{align}

\noindent the subscript $12$ is omitted in this section. Expand $\psi$ and $\Phi$ as
\begin{align}
\psi &= \sum_{l=0}^{\infty} \sum_{m=-l}^l r^l f_{lm}(r) Y_{lm} (\theta, \phi)  \label{y1} \\
\Phi &= \sum_{l=0}^{\infty} \sum_{m=-l}^l r^l \phi_{lm}(r) Y_{lm} (\theta, \phi)  \label{y2} \\
f_{lm}  &= \sum_{k=0}^{\infty} f_{lm}^{(k)} r^k  \\
\phi_{lm} &=  \sum_{k=0}^{\infty} \phi_{lm}^{(k)} r^k
\end{align}

We can establish
\begin{equation}
\sum_{k=0}^{\infty} -\frac{1}{2\mu} k (k+2l+1) f_{lm}^{(k)} r^{l+k-2}   + \sum_{k=0}^{\infty} r^{l+k-1} \phi_{lm}^{(k)} +  O(r^0) \sum_{k=0}^{\infty}  f_{lm}^{(k)} r^{l+k} + O(r^0) \sum_{k=0}^{\infty}  \phi_{lm}^{(k)} r^{l+k} =0
\end{equation}

\noindent from each component of spherical harmonics functions in (\ref{y1}) - (\ref{y2}). Hence
\begin{equation}
 -\frac{1}{2\mu} k (k+2l+1) f_{lm}^{(k)}  +  \phi_{lm}^{(k-1)} +  O(r^0)    f_{lm}^{(k-2)}  + O(r^0)   \phi_{lm}^{(k-2)}  =0
\end{equation}

\noindent For $k=1$ we have obtained the cusp condition
\begin{align}
    f_{lm}^{(1)} =  \frac{ \mu}{l+1} \phi_{lm}^{(0)}
\end{align}

\section{Details of the evaluations for the prefactor of $L^{-8}$ without the presence of geminal functions}

The evaluations of prefactors of $L^{-8}$ are a straightforward continuation of the derivations in literature. We will omit a few definitions and equations which were used by previous authors \cite{Schwartz:63,Kutzelnigg:92}. We start from Eq. (77) in Ref. \cite{Schwartz:63}
\begin{align}
&\left[ s^2 \frac{d^2}{ds^2} - 2y \frac{d}{dy} s \frac{d}{ds}  + y^2 \frac{d^2}{dy^2} + 2y \frac{d}{dy} - 2 s^2 \frac{d}{ds} + 2sy \frac{d}{dy} - \frac{4}{1-y^2/\lambda} \left( 2y \frac{d}{dy} - s \frac{d}{ds} \right)  + \lambda \frac{d^2}{dy^2}  \right. \nonumber \\
&\left. - 4 \left( \lambda -1/4 \right) \frac{ 1+ y^2/\lambda }{ (1-y^2/\lambda)^2}  \right] f_l = \frac{ 2 s e^{-2y} \exp \left[ - (2y^3/3 \lambda) - (2y^5/5\lambda^2) - (2y^7/7\lambda^3)  + \cdots \right] }{   (1-y^2/\lambda)^{1/2} }
\end{align}

Expanding $f_l$ into inverse power series of $\lambda$ and comparing with both sides of Eq. (B.1) will lead to expressions of $f_l$. $f^{(-1)}$ and $f^{(-2)}$ have been published. We obtain next next leading term $f^{(-3)}$
\begin{align}
f^{(-3)} &= 1/11520  e^{-2 y} s \nonumber \\
& ( -2880 + 4275 s - 1440 s^2 - 5760 y + 8550 s y - 2880 s^2 y    \nonumber \\
& -  4320 y^2 + 6300 s y^2 - 2880 s^2 y^2 - 960 y^3 + 1200 s y^3   \nonumber \\
& -  1920 s^2 y^3 + 1320 y^4 - 1200 s y^4 - 960 s^2 y^4 + 1872 y^5   \nonumber \\
&  - 960 s y^5 - 384 s^2 y^5 + 3584 y^6 - 1280 s y^6 - 1280 y^7)  \label{f3}
\end{align}

Inserting the expressions of $f^{(-1)}$ - $f^{(-3)}$ into Eq. (83) in Ref. \cite{Schwartz:63} yields Eq. (\ref{s8}) in the present work.

For the second approach, the second-order energy of the ground state helium atom using the notations in Ref. \cite{Kutzelnigg:92} can be written as
\begin{align}
E_l^{(2)} &=  B_l + D_l + \tilde{E}_l^{(2)} \nonumber \\
B_l &:= \frac{1}{2} \langle \Phi | R_{1l} R_{-1l} P_l^2(\cos \theta_{12}) | \Phi \rangle  \\
D_l &:= \frac{1}{2} \langle \Phi | R_{1l} ( U_{12} )_l P_l^2(\cos \theta_{12}) | \Phi \rangle  \\
 \tilde{E}_l^{(2)} &:= \langle \Phi| (U_{12})_l P_l^2(\cos \theta_{12}) | \chi_l  \rangle
\end{align}
\noindent  where the quantity $\chi_l$ was introduced as
\begin{equation}
\psi  = \frac{1}{2} r_{12} \Phi + \sum_l \chi_l P_l (\cos \theta_{12})
\end{equation}

Hence,
\begin{align}
B_l &= - \frac{ 45}{256} (l+1/2)^{-4} - \frac{225}{2048} ( l+1/2)^{-6} - \frac{1575}{8192} ( l+1/2)^{-8} +\cdots  \\
D_l &= \frac{675}{2048} (l+1/2)^{-6} + \frac{315}{1024} ( l+1/2)^{-8}  +\cdots
\end{align}

$ \tilde{E}_l^{(2)}$ comes from expanding $\frac{1}{9s} r_{12}^3 \Phi, -\frac{1}{6} s\eta^2 r_{12} \Phi, \dots$ in $\chi_l$. The rates of convergence are given in the main text of the present work.


\begin{thebibliography}{58}%
\makeatletter
\providecommand \@ifxundefined [1]{%
 \@ifx{#1\undefined}
}%
\providecommand \@ifnum [1]{%
 \ifnum #1\expandafter \@firstoftwo
 \else \expandafter \@secondoftwo
 \fi
}%
\providecommand \@ifx [1]{%
 \ifx #1\expandafter \@firstoftwo
 \else \expandafter \@secondoftwo
 \fi
}%
\providecommand \natexlab [1]{#1}%
\providecommand \enquote  [1]{``#1''}%
\providecommand \bibnamefont  [1]{#1}%
\providecommand \bibfnamefont [1]{#1}%
\providecommand \citenamefont [1]{#1}%
\providecommand \href@noop [0]{\@secondoftwo}%
\providecommand \href [0]{\begingroup \@sanitize@url \@href}%
\providecommand \@href[1]{\@@startlink{#1}\@@href}%
\providecommand \@@href[1]{\endgroup#1\@@endlink}%
\providecommand \@sanitize@url [0]{\catcode `\\12\catcode `\$12\catcode
  `\&12\catcode `\#12\catcode `\^12\catcode `\_12\catcode `\%12\relax}%
\providecommand \@@startlink[1]{}%
\providecommand \@@endlink[0]{}%
\providecommand \url  [0]{\begingroup\@sanitize@url \@url }%
\providecommand \@url [1]{\endgroup\@href {#1}{\urlprefix }}%
\providecommand \urlprefix  [0]{URL }%
\providecommand \Eprint [0]{\href }%
\providecommand \doibase [0]{http://dx.doi.org/}%
\providecommand \selectlanguage [0]{\@gobble}%
\providecommand \bibinfo  [0]{\@secondoftwo}%
\providecommand \bibfield  [0]{\@secondoftwo}%
\providecommand \translation [1]{[#1]}%
\providecommand \BibitemOpen [0]{}%
\providecommand \bibitemStop [0]{}%
\providecommand \bibitemNoStop [0]{.\EOS\space}%
\providecommand \EOS [0]{\spacefactor3000\relax}%
\providecommand \BibitemShut  [1]{\csname bibitem#1\endcsname}%
\let\auto@bib@innerbib\@empty
\bibitem [{\citenamefont {{Attempts have been made to unify these two aspects.
  See H. Nakatsuji}}(2012)}]{Nakatsuji:12a}%
  \BibitemOpen
  \bibfield  {author} {\bibinfo {author} {\bibnamefont {{Attempts have been
  made to unify these two aspects. See H. Nakatsuji}}},\ }\href@noop {}
  {\bibfield  {journal} {\bibinfo  {journal} {Acc. Chem. Res.}\ }\textbf
  {\bibinfo {volume} {45}},\ \bibinfo {pages} {1480} (\bibinfo {year}
  {2012})}\BibitemShut {NoStop}%
\bibitem [{\citenamefont {Feynman}(1982)}]{Feynman:82}%
  \BibitemOpen
  \bibfield  {author} {\bibinfo {author} {\bibfnamefont {R.}~\bibnamefont
  {Feynman}},\ }\href@noop {} {\bibfield  {journal} {\bibinfo  {journal} {Int.
  J. Theor. Phys.}\ }\textbf {\bibinfo {volume} {21}},\ \bibinfo {pages} {467}
  (\bibinfo {year} {1982})}\BibitemShut {NoStop}%
\bibitem [{\citenamefont {Aspuru-Guzik}\ \emph {et~al.}(2005)\citenamefont
  {Aspuru-Guzik}, \citenamefont {Dutoi}, \citenamefont {Love},\ and\
  \citenamefont {{Head-Gordon}}}]{Alan:05}%
  \BibitemOpen
  \bibfield  {author} {\bibinfo {author} {\bibfnamefont {A.}~\bibnamefont
  {Aspuru-Guzik}}, \bibinfo {author} {\bibfnamefont {A.~D.}\ \bibnamefont
  {Dutoi}}, \bibinfo {author} {\bibfnamefont {P.~J.}\ \bibnamefont {Love}}, \
  and\ \bibinfo {author} {\bibfnamefont {M.}~\bibnamefont {{Head-Gordon}}},\
  }\href@noop {} {\bibfield  {journal} {\bibinfo  {journal} {Science}\ }\textbf
  {\bibinfo {volume} {309}},\ \bibinfo {pages} {1704} (\bibinfo {year}
  {2005})}\BibitemShut {NoStop}%
\bibitem [{\citenamefont {Wecker}\ \emph {et~al.}(2014)\citenamefont {Wecker},
  \citenamefont {Bauer}, \citenamefont {Clark}, \citenamefont {Hastings},\ and\
  \citenamefont {Troyer}}]{Troyer:14}%
  \BibitemOpen
  \bibfield  {author} {\bibinfo {author} {\bibfnamefont {D.}~\bibnamefont
  {Wecker}}, \bibinfo {author} {\bibfnamefont {B.}~\bibnamefont {Bauer}},
  \bibinfo {author} {\bibfnamefont {B.~K.}\ \bibnamefont {Clark}}, \bibinfo
  {author} {\bibfnamefont {M.~B.}\ \bibnamefont {Hastings}}, \ and\ \bibinfo
  {author} {\bibfnamefont {M.}~\bibnamefont {Troyer}},\ }\href@noop {}
  {\bibfield  {journal} {\bibinfo  {journal} {Phys. Rev. A}\ }\textbf {\bibinfo
  {volume} {90}},\ \bibinfo {pages} {022305} (\bibinfo {year}
  {2014})}\BibitemShut {NoStop}%
\bibitem [{\citenamefont {Coester}(1958)}]{Coester:58}%
  \BibitemOpen
  \bibfield  {author} {\bibinfo {author} {\bibfnamefont {F.}~\bibnamefont
  {Coester}},\ }\href@noop {} {\bibfield  {journal} {\bibinfo  {journal} {Nucl.
  Phys.}\ }\textbf {\bibinfo {volume} {7}},\ \bibinfo {pages} {421} (\bibinfo
  {year} {1958})}\BibitemShut {NoStop}%
\bibitem [{\citenamefont {Coester}\ and\ \citenamefont
  {K{\"u}mmel}(1958)}]{Coester:60}%
  \BibitemOpen
  \bibfield  {author} {\bibinfo {author} {\bibfnamefont {F.}~\bibnamefont
  {Coester}}\ and\ \bibinfo {author} {\bibfnamefont {H.}~\bibnamefont
  {K{\"u}mmel}},\ }\href@noop {} {\bibfield  {journal} {\bibinfo  {journal}
  {Nucl. Phys.}\ }\textbf {\bibinfo {volume} {17}},\ \bibinfo {pages} {477}
  (\bibinfo {year} {1958})}\BibitemShut {NoStop}%
\bibitem [{\citenamefont {Bartlett}\ and\ \citenamefont
  {Musia{\l}}(2007)}]{Bartlett:07}%
  \BibitemOpen
  \bibfield  {author} {\bibinfo {author} {\bibfnamefont {R.~J.}\ \bibnamefont
  {Bartlett}}\ and\ \bibinfo {author} {\bibfnamefont {M.}~\bibnamefont
  {Musia{\l}}},\ }\href@noop {} {\bibfield  {journal} {\bibinfo  {journal}
  {Rev. Mod. Phys.}\ }\textbf {\bibinfo {volume} {79}},\ \bibinfo {pages} {291}
  (\bibinfo {year} {2007})}\BibitemShut {NoStop}%
\bibitem [{\citenamefont {Lyakh}\ \emph {et~al.}(2012)\citenamefont {Lyakh},
  \citenamefont {Musia{\l}}, \citenamefont {Lotrich},\ and\ \citenamefont
  {Bartlett}}]{lyakh2012multireference}%
  \BibitemOpen
  \bibfield  {author} {\bibinfo {author} {\bibfnamefont {D.~I.}\ \bibnamefont
  {Lyakh}}, \bibinfo {author} {\bibfnamefont {M.}~\bibnamefont {Musia{\l}}},
  \bibinfo {author} {\bibfnamefont {V.~F.}\ \bibnamefont {Lotrich}}, \ and\
  \bibinfo {author} {\bibfnamefont {R.~J.}\ \bibnamefont {Bartlett}},\
  }\href@noop {} {\bibfield  {journal} {\bibinfo  {journal} {Chem. Rev.}\
  }\textbf {\bibinfo {volume} {112}},\ \bibinfo {pages} {182} (\bibinfo {year}
  {2012})}\BibitemShut {NoStop}%
\bibitem [{\citenamefont {White}(1992)}]{White:92}%
  \BibitemOpen
  \bibfield  {author} {\bibinfo {author} {\bibfnamefont {S.~R.}\ \bibnamefont
  {White}},\ }\href@noop {} {\bibfield  {journal} {\bibinfo  {journal} {Phys.
  Rev. Lett.}\ }\textbf {\bibinfo {volume} {69}},\ \bibinfo {pages} {2863}
  (\bibinfo {year} {1992})}\BibitemShut {NoStop}%
\bibitem [{\citenamefont {Chan}\ and\ \citenamefont
  {{Head-Gordon}}(2003)}]{Head-Gordon:03}%
  \BibitemOpen
  \bibfield  {author} {\bibinfo {author} {\bibfnamefont {G.~K.}\ \bibnamefont
  {Chan}}\ and\ \bibinfo {author} {\bibfnamefont {M.}~\bibnamefont
  {{Head-Gordon}}},\ }\href@noop {} {\bibfield  {journal} {\bibinfo  {journal}
  {J. Chem. Phys.}\ }\textbf {\bibinfo {volume} {118}},\ \bibinfo {pages}
  {8551} (\bibinfo {year} {2003})}\BibitemShut {NoStop}%
\bibitem [{\citenamefont {Kato}(1957)}]{Kato:57}%
  \BibitemOpen
  \bibfield  {author} {\bibinfo {author} {\bibfnamefont {T.}~\bibnamefont
  {Kato}},\ }\href@noop {} {\bibfield  {journal} {\bibinfo  {journal} {Comm.
  Pure Appl. Math.}\ }\textbf {\bibinfo {volume} {10}},\ \bibinfo {pages} {151}
  (\bibinfo {year} {1957})}\BibitemShut {NoStop}%
\bibitem [{\citenamefont {Gilbert}(1963)}]{Gilbert:63}%
  \BibitemOpen
  \bibfield  {author} {\bibinfo {author} {\bibfnamefont {T.~L.}\ \bibnamefont
  {Gilbert}},\ }\href@noop {} {\bibfield  {journal} {\bibinfo  {journal} {Rev.
  Mod. Phys.}\ }\textbf {\bibinfo {volume} {35}},\ \bibinfo {pages} {491}
  (\bibinfo {year} {1963})}\BibitemShut {NoStop}%
\bibitem [{\citenamefont {Prendergast}\ \emph {et~al.}(2001)\citenamefont
  {Prendergast}, \citenamefont {Nolan}, \citenamefont {Filippi}, \citenamefont
  {Fahy},\ and\ \citenamefont {Greer}}]{Geer:01}%
  \BibitemOpen
  \bibfield  {author} {\bibinfo {author} {\bibfnamefont {D.}~\bibnamefont
  {Prendergast}}, \bibinfo {author} {\bibfnamefont {M.}~\bibnamefont {Nolan}},
  \bibinfo {author} {\bibfnamefont {C.}~\bibnamefont {Filippi}}, \bibinfo
  {author} {\bibfnamefont {S.}~\bibnamefont {Fahy}}, \ and\ \bibinfo {author}
  {\bibfnamefont {J.~C.}\ \bibnamefont {Greer}},\ }\href@noop {} {\bibfield
  {journal} {\bibinfo  {journal} {J. Chem. Phys.}\ }\textbf {\bibinfo {volume}
  {115}},\ \bibinfo {pages} {1626} (\bibinfo {year} {2001})}\BibitemShut
  {NoStop}%
\bibitem [{\citenamefont {Fournais}\ \emph {et~al.}(2005)\citenamefont
  {Fournais}, \citenamefont {{Hoffmann-Ostenhof}}, \citenamefont
  {{Hoffmann-Otenhof}},\ and\ \citenamefont {S{\o}ensen}}]{Fournais:05}%
  \BibitemOpen
  \bibfield  {author} {\bibinfo {author} {\bibfnamefont {S.}~\bibnamefont
  {Fournais}}, \bibinfo {author} {\bibfnamefont {M.}~\bibnamefont
  {{Hoffmann-Ostenhof}}}, \bibinfo {author} {\bibfnamefont {T.}~\bibnamefont
  {{Hoffmann-Otenhof}}}, \ and\ \bibinfo {author} {\bibfnamefont {T.~{\O}.}\
  \bibnamefont {S{\o}ensen}},\ }\href@noop {} {\bibfield  {journal} {\bibinfo
  {journal} {Commun. Math. Phys.}\ }\textbf {\bibinfo {volume} {255}},\
  \bibinfo {pages} {183} (\bibinfo {year} {2005})}\BibitemShut {NoStop}%
\bibitem [{\citenamefont {Nakashima}\ and\ \citenamefont
  {Nakatsuji}(2007)}]{Nakatsuji:07}%
  \BibitemOpen
  \bibfield  {author} {\bibinfo {author} {\bibfnamefont {H.}~\bibnamefont
  {Nakashima}}\ and\ \bibinfo {author} {\bibfnamefont {H.}~\bibnamefont
  {Nakatsuji}},\ }\href@noop {} {\bibfield  {journal} {\bibinfo  {journal} {J.
  Chem. Phys.}\ }\textbf {\bibinfo {volume} {127}},\ \bibinfo {pages} {224104}
  (\bibinfo {year} {2007})}\BibitemShut {NoStop}%
\bibitem [{\citenamefont {Fournais}\ \emph {et~al.}(2009)\citenamefont
  {Fournais}, \citenamefont {{Hoffmann-Ostenhof}}, \citenamefont
  {{Hoffmann-Otenhof}},\ and\ \citenamefont {S{\o}ensen}}]{Fournais:09}%
  \BibitemOpen
  \bibfield  {author} {\bibinfo {author} {\bibfnamefont {S.}~\bibnamefont
  {Fournais}}, \bibinfo {author} {\bibfnamefont {M.}~\bibnamefont
  {{Hoffmann-Ostenhof}}}, \bibinfo {author} {\bibfnamefont {T.}~\bibnamefont
  {{Hoffmann-Otenhof}}}, \ and\ \bibinfo {author} {\bibfnamefont {T.~{\O}.}\
  \bibnamefont {S{\o}ensen}},\ }\href@noop {} {\bibfield  {journal} {\bibinfo
  {journal} {Commun. Math. Phys.}\ }\textbf {\bibinfo {volume} {94}},\ \bibinfo
  {pages} {289} (\bibinfo {year} {2009})}\BibitemShut {NoStop}%
\bibitem [{\citenamefont {Kutzelnigg}\ and\ \citenamefont {{Morgan
  III}}(1992)}]{Kutzelnigg:92}%
  \BibitemOpen
  \bibfield  {author} {\bibinfo {author} {\bibfnamefont {W.}~\bibnamefont
  {Kutzelnigg}}\ and\ \bibinfo {author} {\bibfnamefont {J.~D.}\ \bibnamefont
  {{Morgan III}}},\ }\href@noop {} {\bibfield  {journal} {\bibinfo  {journal}
  {J. Chem. Phys.}\ }\textbf {\bibinfo {volume} {96}},\ \bibinfo {pages} {4484}
  (\bibinfo {year} {1992})}\BibitemShut {NoStop}%
\bibitem [{\citenamefont {Schwartz}(1962)}]{Schwartz:62}%
  \BibitemOpen
  \bibfield  {author} {\bibinfo {author} {\bibfnamefont {C.}~\bibnamefont
  {Schwartz}},\ }\href@noop {} {\bibfield  {journal} {\bibinfo  {journal}
  {Phys. Rev.}\ }\textbf {\bibinfo {volume} {126}},\ \bibinfo {pages} {1015}
  (\bibinfo {year} {1962})}\BibitemShut {NoStop}%
\bibitem [{\citenamefont {Schwartz}(1963)}]{Schwartz:63}%
  \BibitemOpen
  \bibfield  {author} {\bibinfo {author} {\bibfnamefont {C.}~\bibnamefont
  {Schwartz}},\ }\href@noop {} {\bibfield  {journal} {\bibinfo  {journal}
  {Methods Comput. Phys.}\ }\textbf {\bibinfo {volume} {2}},\ \bibinfo {pages}
  {241} (\bibinfo {year} {1963})}\BibitemShut {NoStop}%
\bibitem [{\citenamefont {Hill}(1985)}]{Hill:85}%
  \BibitemOpen
  \bibfield  {author} {\bibinfo {author} {\bibfnamefont {R.~N.}\ \bibnamefont
  {Hill}},\ }\href@noop {} {\bibfield  {journal} {\bibinfo  {journal} {J. Chem.
  Phys.}\ }\textbf {\bibinfo {volume} {83}},\ \bibinfo {pages} {1173} (\bibinfo
  {year} {1985})}\BibitemShut {NoStop}%
\bibitem [{\citenamefont {Goddard}(2009)}]{Goddard:09b}%
  \BibitemOpen
  \bibfield  {author} {\bibinfo {author} {\bibfnamefont {B.~D.}\ \bibnamefont
  {Goddard}},\ }\href@noop {} {\bibfield  {journal} {\bibinfo  {journal} {Siam
  J. Math. Anal.}\ }\textbf {\bibinfo {volume} {41}},\ \bibinfo {pages} {77}
  (\bibinfo {year} {2009})}\BibitemShut {NoStop}%
\bibitem [{\citenamefont {Halkier}\ \emph {et~al.}(1999)\citenamefont
  {Halkier}, \citenamefont {Helgaker}, \citenamefont {J{\o}rgensen},
  \citenamefont {Klopper},\ and\ \citenamefont {Olsen}}]{Klopper:99}%
  \BibitemOpen
  \bibfield  {author} {\bibinfo {author} {\bibfnamefont {A.}~\bibnamefont
  {Halkier}}, \bibinfo {author} {\bibfnamefont {T.}~\bibnamefont {Helgaker}},
  \bibinfo {author} {\bibfnamefont {P.}~\bibnamefont {J{\o}rgensen}}, \bibinfo
  {author} {\bibfnamefont {W.}~\bibnamefont {Klopper}}, \ and\ \bibinfo
  {author} {\bibfnamefont {J.}~\bibnamefont {Olsen}},\ }\href@noop {}
  {\bibfield  {journal} {\bibinfo  {journal} {Chem. Phys. Lett.}\ }\textbf
  {\bibinfo {volume} {302}},\ \bibinfo {pages} {437} (\bibinfo {year}
  {1999})}\BibitemShut {NoStop}%
\bibitem [{\citenamefont {Jankowski}\ \emph {et~al.}(2006)\citenamefont
  {Jankowski}, \citenamefont {S{\l}upski},\ and\ \citenamefont
  {Flores}}]{Jankowski:06}%
  \BibitemOpen
  \bibfield  {author} {\bibinfo {author} {\bibfnamefont {K.}~\bibnamefont
  {Jankowski}}, \bibinfo {author} {\bibfnamefont {R.}~\bibnamefont
  {S{\l}upski}}, \ and\ \bibinfo {author} {\bibfnamefont {J.~R.}\ \bibnamefont
  {Flores}},\ }\href@noop {} {\bibfield  {journal} {\bibinfo  {journal} {Mol.
  Phys.}\ }\textbf {\bibinfo {volume} {104}},\ \bibinfo {pages} {2213}
  (\bibinfo {year} {2006})}\BibitemShut {NoStop}%
\bibitem [{\citenamefont {H{\"a}ttig}\ \emph {et~al.}(2012)\citenamefont
  {H{\"a}ttig}, \citenamefont {Klopper}, \citenamefont {K{\"o}hn},\ and\
  \citenamefont {Tew}}]{Klopper:12}%
  \BibitemOpen
  \bibfield  {author} {\bibinfo {author} {\bibfnamefont {C.}~\bibnamefont
  {H{\"a}ttig}}, \bibinfo {author} {\bibfnamefont {W.}~\bibnamefont {Klopper}},
  \bibinfo {author} {\bibfnamefont {A.}~\bibnamefont {K{\"o}hn}}, \ and\
  \bibinfo {author} {\bibfnamefont {D.~P.}\ \bibnamefont {Tew}},\ }\href@noop
  {} {\bibfield  {journal} {\bibinfo  {journal} {Chem. Rev.}\ }\textbf
  {\bibinfo {volume} {112}},\ \bibinfo {pages} {4} (\bibinfo {year}
  {2012})}\BibitemShut {NoStop}%
\bibitem [{\citenamefont {Kong}\ \emph {et~al.}(2012)\citenamefont {Kong},
  \citenamefont {Bischoff},\ and\ \citenamefont {Valeev}}]{Valeev:12}%
  \BibitemOpen
  \bibfield  {author} {\bibinfo {author} {\bibfnamefont {L.}~\bibnamefont
  {Kong}}, \bibinfo {author} {\bibfnamefont {F.~A.}\ \bibnamefont {Bischoff}},
  \ and\ \bibinfo {author} {\bibfnamefont {E.~F.}\ \bibnamefont {Valeev}},\
  }\href@noop {} {\bibfield  {journal} {\bibinfo  {journal} {Chem. Rev.}\
  }\textbf {\bibinfo {volume} {112}},\ \bibinfo {pages} {75} (\bibinfo {year}
  {2012})}\BibitemShut {NoStop}%
\bibitem [{\citenamefont {Ten-no}\ and\ \citenamefont
  {Noga}(2012)}]{Ten-no:12}%
  \BibitemOpen
  \bibfield  {author} {\bibinfo {author} {\bibfnamefont {S.}~\bibnamefont
  {Ten-no}}\ and\ \bibinfo {author} {\bibfnamefont {J.}~\bibnamefont {Noga}},\
  }\href@noop {} {\bibfield  {journal} {\bibinfo  {journal} {Wiley Interdiscip.
  Rev. Comput. Mol. Sci.}\ }\textbf {\bibinfo {volume} {2}},\ \bibinfo {pages}
  {114} (\bibinfo {year} {2012})}\BibitemShut {NoStop}%
\bibitem [{\citenamefont {Hirata}(2012)}]{Hirata:12}%
  \BibitemOpen
  \bibfield  {author} {\bibinfo {author} {\bibfnamefont {S.}~\bibnamefont
  {Hirata}},\ }\href@noop {} {\bibfield  {journal} {\bibinfo  {journal} {Theor.
  Chim. Acta.}\ }\textbf {\bibinfo {volume} {131}},\ \bibinfo {pages} {1071}
  (\bibinfo {year} {2012})}\BibitemShut {NoStop}%
\bibitem [{\citenamefont {Schwenke}(2012)}]{Schwenke:12}%
  \BibitemOpen
  \bibfield  {author} {\bibinfo {author} {\bibfnamefont {D.~W.}\ \bibnamefont
  {Schwenke}},\ }\href@noop {} {\bibfield  {journal} {\bibinfo  {journal} {Mol.
  Phys.}\ }\textbf {\bibinfo {volume} {110}},\ \bibinfo {pages} {2557}
  (\bibinfo {year} {2012})}\BibitemShut {NoStop}%
\bibitem [{\citenamefont {Shepherd}\ \emph {et~al.}(2012)\citenamefont
  {Shepherd}, \citenamefont {Gr{\"u}neis}, \citenamefont {Booth}, \citenamefont
  {Kresse},\ and\ \citenamefont {Alavi}}]{Shepherd:12}%
  \BibitemOpen
  \bibfield  {author} {\bibinfo {author} {\bibfnamefont {J.~J.}\ \bibnamefont
  {Shepherd}}, \bibinfo {author} {\bibfnamefont {A.}~\bibnamefont
  {Gr{\"u}neis}}, \bibinfo {author} {\bibfnamefont {G.~H.}\ \bibnamefont
  {Booth}}, \bibinfo {author} {\bibfnamefont {G.}~\bibnamefont {Kresse}}, \
  and\ \bibinfo {author} {\bibfnamefont {A.}~\bibnamefont {Alavi}},\
  }\href@noop {} {\bibfield  {journal} {\bibinfo  {journal} {Phys. Rev. B}\
  }\textbf {\bibinfo {volume} {86}},\ \bibinfo {pages} {035111} (\bibinfo
  {year} {2012})}\BibitemShut {NoStop}%
\bibitem [{\citenamefont {Ten-no}(2004)}]{Ten-no:04}%
  \BibitemOpen
  \bibfield  {author} {\bibinfo {author} {\bibfnamefont {S.}~\bibnamefont
  {Ten-no}},\ }\href@noop {} {\bibfield  {journal} {\bibinfo  {journal} {Chem.
  Phys. Lett.}\ }\textbf {\bibinfo {volume} {398}},\ \bibinfo {pages} {56}
  (\bibinfo {year} {2004})}\BibitemShut {NoStop}%
\bibitem [{\citenamefont {Lesiuk}\ \emph {et~al.}(2013)\citenamefont {Lesiuk},
  \citenamefont {Jeziorski},\ and\ \citenamefont {Moszynski}}]{Jeziorski:13}%
  \BibitemOpen
  \bibfield  {author} {\bibinfo {author} {\bibfnamefont {M.}~\bibnamefont
  {Lesiuk}}, \bibinfo {author} {\bibfnamefont {B.}~\bibnamefont {Jeziorski}}, \
  and\ \bibinfo {author} {\bibfnamefont {R.}~\bibnamefont {Moszynski}},\
  }\href@noop {} {\bibfield  {journal} {\bibinfo  {journal} {J. Chem. Phys.}\
  }\textbf {\bibinfo {volume} {139}},\ \bibinfo {pages} {134102} (\bibinfo
  {year} {2013})}\BibitemShut {NoStop}%
\bibitem [{\citenamefont {Silkowski}\ \emph {et~al.}(2015)\citenamefont
  {Silkowski}, \citenamefont {Lesiuk},\ and\ \citenamefont
  {Moszynski}}]{:/content/aip/journal/jcp/142/12/10.1063/1.4915272}%
  \BibitemOpen
  \bibfield  {author} {\bibinfo {author} {\bibfnamefont {M.}~\bibnamefont
  {Silkowski}}, \bibinfo {author} {\bibfnamefont {M.}~\bibnamefont {Lesiuk}}, \
  and\ \bibinfo {author} {\bibfnamefont {R.}~\bibnamefont {Moszynski}},\ }\href
  {\doibase http://dx.doi.org/10.1063/1.4915272} {\bibfield  {journal}
  {\bibinfo  {journal} {J. Chem. Phys.}\ }\textbf {\bibinfo {volume} {142}},\
  \bibinfo {eid} {124102} (\bibinfo {year} {2015})}\BibitemShut {NoStop}%
\bibitem [{\citenamefont {K{\"o}hn}(2010)}]{Kohn:10}%
  \BibitemOpen
  \bibfield  {author} {\bibinfo {author} {\bibfnamefont {A.}~\bibnamefont
  {K{\"o}hn}},\ }\href@noop {} {\bibfield  {journal} {\bibinfo  {journal} {J.
  Chem. Phys.}\ }\textbf {\bibinfo {volume} {133}},\ \bibinfo {pages} {174118}
  (\bibinfo {year} {2010})}\BibitemShut {NoStop}%
\bibitem [{\citenamefont {Shiozaki}\ \emph {et~al.}(2008)\citenamefont
  {Shiozaki}, \citenamefont {Kamiya}, \citenamefont {Hirata},\ and\
  \citenamefont {Valeev}}]{Hirata:08}%
  \BibitemOpen
  \bibfield  {author} {\bibinfo {author} {\bibfnamefont {T.}~\bibnamefont
  {Shiozaki}}, \bibinfo {author} {\bibfnamefont {M.}~\bibnamefont {Kamiya}},
  \bibinfo {author} {\bibfnamefont {S.}~\bibnamefont {Hirata}}, \ and\ \bibinfo
  {author} {\bibfnamefont {E.~F.}\ \bibnamefont {Valeev}},\ }\href@noop {}
  {\bibfield  {journal} {\bibinfo  {journal} {J. Chem. Phys.}\ }\textbf
  {\bibinfo {volume} {130}},\ \bibinfo {pages} {054101} (\bibinfo {year}
  {2008})}\BibitemShut {NoStop}%
\bibitem [{\citenamefont {Yanai}\ and\ \citenamefont
  {Shiozaki}(2012)}]{Yanai:12}%
  \BibitemOpen
  \bibfield  {author} {\bibinfo {author} {\bibfnamefont {T.}~\bibnamefont
  {Yanai}}\ and\ \bibinfo {author} {\bibfnamefont {T.}~\bibnamefont
  {Shiozaki}},\ }\href@noop {} {\bibfield  {journal} {\bibinfo  {journal} {J.
  Chem. Phys.}\ }\textbf {\bibinfo {volume} {136}},\ \bibinfo {pages} {084107}
  (\bibinfo {year} {2012})}\BibitemShut {NoStop}%
\bibitem [{\citenamefont {Rassolov}\ and\ \citenamefont
  {Chipman}(1996)}]{Rassolov:96}%
  \BibitemOpen
  \bibfield  {author} {\bibinfo {author} {\bibfnamefont {V.~A.}\ \bibnamefont
  {Rassolov}}\ and\ \bibinfo {author} {\bibfnamefont {D.~M.}\ \bibnamefont
  {Chipman}},\ }\href@noop {} {\bibfield  {journal} {\bibinfo  {journal} {J.
  Chem. Phys.}\ }\textbf {\bibinfo {volume} {104}},\ \bibinfo {pages} {9908}
  (\bibinfo {year} {1996})}\BibitemShut {NoStop}%
\bibitem [{\citenamefont {Tew}(2008)}]{Tew:08}%
  \BibitemOpen
  \bibfield  {author} {\bibinfo {author} {\bibfnamefont {D.~P.}\ \bibnamefont
  {Tew}},\ }\href@noop {} {\bibfield  {journal} {\bibinfo  {journal} {J. Chem.
  Phys.}\ }\textbf {\bibinfo {volume} {129}},\ \bibinfo {pages} {014104}
  (\bibinfo {year} {2008})}\BibitemShut {NoStop}%
\bibitem [{\citenamefont {Kurokawa}\ \emph {et~al.}(2013)\citenamefont
  {Kurokawa}, \citenamefont {Nakashima},\ and\ \citenamefont
  {Nakatsuji}}]{Nakatsuji:13}%
  \BibitemOpen
  \bibfield  {author} {\bibinfo {author} {\bibfnamefont {Y.~I.}\ \bibnamefont
  {Kurokawa}}, \bibinfo {author} {\bibfnamefont {H.}~\bibnamefont {Nakashima}},
  \ and\ \bibinfo {author} {\bibfnamefont {H.}~\bibnamefont {Nakatsuji}},\
  }\href@noop {} {\bibfield  {journal} {\bibinfo  {journal} {J. Chem. Phys.}\
  }\textbf {\bibinfo {volume} {139}},\ \bibinfo {pages} {044114} (\bibinfo
  {year} {2013})}\BibitemShut {NoStop}%
\bibitem [{\citenamefont {Kurokawa}\ \emph {et~al.}(2014)\citenamefont
  {Kurokawa}, \citenamefont {Nakashima},\ and\ \citenamefont
  {Nakatsuji}}]{Nakatsuji:14}%
  \BibitemOpen
  \bibfield  {author} {\bibinfo {author} {\bibfnamefont {Y.~I.}\ \bibnamefont
  {Kurokawa}}, \bibinfo {author} {\bibfnamefont {H.}~\bibnamefont {Nakashima}},
  \ and\ \bibinfo {author} {\bibfnamefont {H.}~\bibnamefont {Nakatsuji}},\
  }\href@noop {} {\bibfield  {journal} {\bibinfo  {journal} {J. Chem. Phys.}\
  }\textbf {\bibinfo {volume} {140}},\ \bibinfo {pages} {214103} (\bibinfo
  {year} {2014})}\BibitemShut {NoStop}%
\bibitem [{\citenamefont {Wang}(2013)}]{Wang:13}%
  \BibitemOpen
  \bibfield  {author} {\bibinfo {author} {\bibfnamefont {C.}~\bibnamefont
  {Wang}},\ }\href@noop {} {\bibfield  {journal} {\bibinfo  {journal} {Phys.
  Rev. A}\ }\textbf {\bibinfo {volume} {88}},\ \bibinfo {pages} {032511}
  (\bibinfo {year} {2013})}\BibitemShut {NoStop}%
\bibitem [{\citenamefont {Wang}(2015)}]{Wang:15a}%
  \BibitemOpen
  \bibfield  {author} {\bibinfo {author} {\bibfnamefont {C.}~\bibnamefont
  {Wang}},\ }\href@noop {} {\bibfield  {journal} {\bibinfo  {journal} {Phys.
  Rev. A}\ }\textbf {\bibinfo {volume} {91}},\ \bibinfo {pages} {069904}
  (\bibinfo {year} {2015})}\BibitemShut {NoStop}%
\bibitem [{\citenamefont {White}\ and\ \citenamefont
  {Brown}(1967)}]{White:67a}%
  \BibitemOpen
  \bibfield  {author} {\bibinfo {author} {\bibfnamefont {R.~J.}\ \bibnamefont
  {White}}\ and\ \bibinfo {author} {\bibfnamefont {W.~B.}\ \bibnamefont
  {Brown}},\ }\href@noop {} {\bibfield  {journal} {\bibinfo  {journal} {Int. J
  Quantum Chem.}\ }\textbf {\bibinfo {volume} {1}},\ \bibinfo {pages} {61}
  (\bibinfo {year} {1967})}\BibitemShut {NoStop}%
\bibitem [{\citenamefont {Brown}\ and\ \citenamefont
  {White}(1967{\natexlab{a}})}]{White:67c}%
  \BibitemOpen
  \bibfield  {author} {\bibinfo {author} {\bibfnamefont {W.~B.}\ \bibnamefont
  {Brown}}\ and\ \bibinfo {author} {\bibfnamefont {R.~J.}\ \bibnamefont
  {White}},\ }\href@noop {} {\bibfield  {journal} {\bibinfo  {journal} {Phys.
  Rev. Lett.}\ }\textbf {\bibinfo {volume} {18}},\ \bibinfo {pages} {1178}
  (\bibinfo {year} {1967}{\natexlab{a}})}\BibitemShut {NoStop}%
\bibitem [{\citenamefont {Brown}\ and\ \citenamefont
  {White}(1967{\natexlab{b}})}]{White:67d}%
  \BibitemOpen
  \bibfield  {author} {\bibinfo {author} {\bibfnamefont {W.~B.}\ \bibnamefont
  {Brown}}\ and\ \bibinfo {author} {\bibfnamefont {R.~J.}\ \bibnamefont
  {White}},\ }\href@noop {} {\bibfield  {journal} {\bibinfo  {journal} {Phys.
  Rev. Lett.}\ }\textbf {\bibinfo {volume} {18}},\ \bibinfo {pages} {1037}
  (\bibinfo {year} {1967}{\natexlab{b}})}\BibitemShut {NoStop}%
\bibitem [{\citenamefont {Hylleraas}(1930)}]{Hylleraas:30}%
  \BibitemOpen
  \bibfield  {author} {\bibinfo {author} {\bibfnamefont {E.~A.}\ \bibnamefont
  {Hylleraas}},\ }\href@noop {} {\bibfield  {journal} {\bibinfo  {journal} {Z.
  Phys.}\ }\textbf {\bibinfo {volume} {65}},\ \bibinfo {pages} {209} (\bibinfo
  {year} {1930})}\BibitemShut {NoStop}%
\bibitem [{\citenamefont {Sack}(1964)}]{Sack:64}%
  \BibitemOpen
  \bibfield  {author} {\bibinfo {author} {\bibfnamefont {R.~A.}\ \bibnamefont
  {Sack}},\ }\href@noop {} {\bibfield  {journal} {\bibinfo  {journal} {J. Math.
  Phys.}\ }\textbf {\bibinfo {volume} {5}},\ \bibinfo {pages} {245} (\bibinfo
  {year} {1964})}\BibitemShut {NoStop}%
\bibitem [{\citenamefont {Cohl}(2013)}]{Cohl:13}%
  \BibitemOpen
  \bibfield  {author} {\bibinfo {author} {\bibfnamefont {H.~S.}\ \bibnamefont
  {Cohl}},\ }\href@noop {} {\bibfield  {journal} {\bibinfo  {journal} {Integr.
  Transf. Spec. F.}\ }\textbf {\bibinfo {volume} {24}},\ \bibinfo {pages} {807}
  (\bibinfo {year} {2013})}\BibitemShut {NoStop}%
\bibitem [{\citenamefont {Hameka}(p223)}]{Hameka:67}%
  \BibitemOpen
  \bibfield  {author} {\bibinfo {author} {\bibfnamefont {H.~F.}\ \bibnamefont
  {Hameka}},\ }\href@noop {} {\emph {\bibinfo {title} {Introduction to quantum
  theory}}}\ (\bibinfo  {publisher} {Harper \& Row},\ \bibinfo {address} {New
  York},\ \bibinfo {year} {1967, p.223})\BibitemShut {NoStop}%
\bibitem [{\citenamefont {{Byron Jr.}}\ and\ \citenamefont
  {Joachain}(1966)}]{Byron:67}%
  \BibitemOpen
  \bibfield  {author} {\bibinfo {author} {\bibfnamefont {F.~W.}\ \bibnamefont
  {{Byron Jr.}}}\ and\ \bibinfo {author} {\bibfnamefont {C.~J.}\ \bibnamefont
  {Joachain}},\ }\href@noop {} {\bibfield  {journal} {\bibinfo  {journal}
  {Phys. Rev.}\ }\textbf {\bibinfo {volume} {157}},\ \bibinfo {pages} {1}
  (\bibinfo {year} {1966})}\BibitemShut {NoStop}%
\bibitem [{\citenamefont {White}(1967)}]{White:67b}%
  \BibitemOpen
  \bibfield  {author} {\bibinfo {author} {\bibfnamefont {R.~J.}\ \bibnamefont
  {White}},\ }\href@noop {} {\bibfield  {journal} {\bibinfo  {journal} {Phys.
  Rev.}\ }\textbf {\bibinfo {volume} {154}},\ \bibinfo {pages} {116} (\bibinfo
  {year} {1967})}\BibitemShut {NoStop}%
\bibitem [{\citenamefont {Schmidt}\ and\ \citenamefont
  {v~Hirschhausen}(1983)}]{Schmidt:83}%
  \BibitemOpen
  \bibfield  {author} {\bibinfo {author} {\bibfnamefont {H.~M.}\ \bibnamefont
  {Schmidt}}\ and\ \bibinfo {author} {\bibfnamefont {H.}~\bibnamefont
  {v~Hirschhausen}},\ }\href@noop {} {\bibfield  {journal} {\bibinfo  {journal}
  {Phys. Rev. A}\ }\textbf {\bibinfo {volume} {28}},\ \bibinfo {pages} {3179}
  (\bibinfo {year} {1983})}\BibitemShut {NoStop}%
\bibitem [{\citenamefont {Helgaker}\ \emph {et~al.}(2000)\citenamefont
  {Helgaker}, \citenamefont {J{\o}rgensen},\ and\ \citenamefont
  {Olsen}}]{Helgaker:2000}%
  \BibitemOpen
  \bibfield  {author} {\bibinfo {author} {\bibfnamefont {T.}~\bibnamefont
  {Helgaker}}, \bibinfo {author} {\bibfnamefont {P.}~\bibnamefont
  {J{\o}rgensen}}, \ and\ \bibinfo {author} {\bibfnamefont {J.}~\bibnamefont
  {Olsen}},\ }\href@noop {} {\emph {\bibinfo {title} {Molecular
  Electronic-Structure Theory}}}\ (\bibinfo  {publisher} {Wiley},\ \bibinfo
  {address} {Chichester},\ \bibinfo {year} {2000})\BibitemShut {NoStop}%
\bibitem [{\citenamefont {Kutzelnigg}(1985)}]{kutzelnigg1985r}%
  \BibitemOpen
  \bibfield  {author} {\bibinfo {author} {\bibfnamefont {W.}~\bibnamefont
  {Kutzelnigg}},\ }\href@noop {} {\bibfield  {journal} {\bibinfo  {journal}
  {Theor. Chim. Acta}\ }\textbf {\bibinfo {volume} {68}},\ \bibinfo {pages}
  {445} (\bibinfo {year} {1985})}\BibitemShut {NoStop}%
\bibitem [{\citenamefont {{Maple 2016. Maplesoft, a division of Waterloo Maple
  Inc., Waterloo, Ontario.}}()}]{MAPLE:11}%
  \BibitemOpen
  \bibfield  {author} {\bibinfo {author} {\bibnamefont {{Maple 2016. Maplesoft,
  a division of Waterloo Maple Inc., Waterloo, Ontario.}}},\ }\href@noop {} {}\
  (\bibinfo {address} {Waterloo, Canada})\BibitemShut {NoStop}%
\bibitem [{\citenamefont {Schwartz}(2006{\natexlab{a}})}]{Schwartz:06a}%
  \BibitemOpen
  \bibfield  {author} {\bibinfo {author} {\bibfnamefont {C.}~\bibnamefont
  {Schwartz}},\ }\href@noop {} {\bibfield  {journal} {\bibinfo  {journal} {Int.
  J. Mod. Phys. E}\ }\textbf {\bibinfo {volume} {28}},\ \bibinfo {pages} {1}
  (\bibinfo {year} {2006}{\natexlab{a}})}\BibitemShut {NoStop}%
\bibitem [{\citenamefont {Schwartz}(2006{\natexlab{b}})}]{Schwartz:06b}%
  \BibitemOpen
  \bibfield  {author} {\bibinfo {author} {\bibfnamefont {C.}~\bibnamefont
  {Schwartz}},\ }\href@noop {} {\bibfield  {journal} {\bibinfo  {journal}
  {arXiv:math-ph/0605018}\ } (\bibinfo {year}
  {2006}{\natexlab{b}})}\BibitemShut {NoStop}%
\bibitem [{\citenamefont {Kurokawa}\ \emph {et~al.}(2008)\citenamefont
  {Kurokawa}, \citenamefont {Nakashima},\ and\ \citenamefont
  {Nakatsuji}}]{kurokawa2008solving}%
  \BibitemOpen
  \bibfield  {author} {\bibinfo {author} {\bibfnamefont {Y.~I.}\ \bibnamefont
  {Kurokawa}}, \bibinfo {author} {\bibfnamefont {H.}~\bibnamefont {Nakashima}},
  \ and\ \bibinfo {author} {\bibfnamefont {H.}~\bibnamefont {Nakatsuji}},\
  }\href@noop {} {\bibfield  {journal} {\bibinfo  {journal} {Phys. Chem. Chem.
  Phys}\ }\textbf {\bibinfo {volume} {10}},\ \bibinfo {pages} {4486} (\bibinfo
  {year} {2008})}\BibitemShut {NoStop}%
\bibitem [{\citenamefont {Pack}\ and\ \citenamefont {Brown}(1966)}]{Pack:66}%
  \BibitemOpen
  \bibfield  {author} {\bibinfo {author} {\bibfnamefont {R.~T.}\ \bibnamefont
  {Pack}}\ and\ \bibinfo {author} {\bibfnamefont {W.~B.}\ \bibnamefont
  {Brown}},\ }\href@noop {} {\bibfield  {journal} {\bibinfo  {journal} {J.
  Chem. Phys.}\ }\textbf {\bibinfo {volume} {45}},\ \bibinfo {pages} {556}
  (\bibinfo {year} {1966})}\BibitemShut {NoStop}%
\end{thebibliography}
\end{document}